\documentclass[cameraready]{Interspeech}
\title{
Something from Nothing: Data Augmentation for\\Robust Severity Level Estimation of Dysarthric Speech}

\author[affiliation={1}, orcid=0009-0004-4581-8364, equalcontribution]{Jaesung}{Bae}
\author[affiliation={1}, orcid=0009-0002-2500-9805, equalcontribution]{Xiuwen}{Zheng}
\author[affiliation={1}, orcid=0000-0003-3513-8328]{Minje}{Kim}
\author[affiliation={2}, orcid=0000-0002-0756-7179]{Chang D.}{Yoo}
\author[affiliation={1}, orcid=0000-0002-5631-2893, correspondingauthor]{Mark}{Hasegawa-Johnson}
\address{
    $^1$ University of Illinois Urbana-Champaign, IL, USA \\
    $^2$ Korea Advanced Institute of Science \& Technology, KR
}

\email{\{jb82, xiuwenz2, minje, jhasegaw\}@illinois.edu, cd\_yoo@kaist.ac.kr
}


\usepackage{comment}

\usepackage{amsmath}
\usepackage{amssymb}
\usepackage{multirow}
\usepackage[table]{xcolor}
\usepackage{graphicx}
\usepackage{subcaption}
\usepackage{amssymb}
\usepackage{pifont}
\newcommand{\cmark}{\ding{51}}  % checkmark
\newcommand{\xmark}{\ding{55}}  % cross
\usepackage{mathtools}
\usepackage{amsmath}
\usepackage{tabularx}
\usepackage{booktabs}
\newcolumntype{Y}{>{\raggedright\arraybackslash}X}

\newif\iftodo
\newif\ifminje
\newif\ifMH
\todotrue
\minjetrue
\MHtrue

\newlength{\vspaceval}
\newlength{\tablecaptionvspaceval}
\begin{document}

\maketitle

\begin{abstract}
Dysarthric speech quality assessment (DSQA) is critical for clinical diagnostics and inclusive speech technologies. However, subjective evaluation is costly and difficult to scale, and the scarcity of labeled data limits robust objective modeling. To address this, we propose a three-stage framework that leverages unlabeled dysarthric speech and large-scale typical speech datasets to scale training. A teacher model first generates pseudo-labels for unlabeled samples, followed by weakly supervised pretraining using a label-aware contrastive learning strategy that exposes the model to diverse speakers and acoustic conditions. The pretrained model is then fine-tuned for the downstream DSQA task. Experiments on five unseen datasets spanning multiple etiologies and languages demonstrate the robustness of our approach. Our Whisper-based baseline significantly outperforms SOTA DSQA predictors such as SpICE, and the full framework achieves an average SRCC of 0.761 across unseen test datasets.

\end{abstract}

\section{Introduction}
\label{sec:intro}

Dysarthria is a motor speech disorder caused by neurological impairments, leading to substantial degradation in acoustic and perceptual characteristics. Accurate dysarthric speech quality assessment (DSQA) is essential for clinical diagnosis, early detection of progressive neurological conditions, rehabilitation monitoring, and the development of inclusive speech technologies, including pathological speech enhancement and automatic speech recognition. However, DSQA relies on expert clinical evaluation by speech-language pathologists (SLPs), limiting its scalability and accessibility in real-world settings. An automated and objective DSQA system would therefore complement SLP expertise by enabling continuous monitoring outside the clinic.

Recent deep learning-based SQA models can reliably assess the perceptual correlates of algorithmic and channel distortion; in particular, non-intrusive SQA (NI-SQA) approaches estimate speech quality directly from the input signal without reference recordings. For example, DNSMOS~\cite{reddy2021dnsmos} evaluates speech quality in real-world noisy and reverberant conditions, while UTMOS~\cite{saeki2022utmos} predicts subjective quality for clean and enhanced speech. The growing availability of large-scale dysarthric datasets has further enabled the development of deep learning-based DSQA models. SpICE~\cite{venugopalan2023spice} is trained on 550,000 disordered speech samples from \textit{Project Euphonia}~\cite{macdonald2021disordered} to automatically predict human-perceived intelligibility. More recently,~\cite{narain2025voicequality} develops voice quality probes for dysarthric speech across seven perceptual dimensions, trained on 11,184 samples from the \textit{Speech Accessibility Project (SAP)}~\cite{hasegawa2024community}. These previous approaches show promising results; however, most evaluations have been conducted on English corpora, which may limit their generalizability to non-English datasets.

Contrastive learning methods such as SimCLR~\cite{simclr} are a powerful self-supervised learning method that can form a structured latent representation space. It has also been widely adopted in speech technology, e.g., to train speech foundation models such as wav2vec 2.0~\cite{baevski2020wav2vec} and HuBERT~\cite{hsu2021hubert}. This representation space is also proven to be effective for DSQA approaches, such as in ~\cite{venugopalan2023spice,narain2025voicequality}. However, to transfer the speech foundation model into the downstream DSQA task, a labeled dataset is required, and the availability of in-domain training data remains scarce. The Euphonia dataset~\cite{macdonald2021disordered} is not publicly released, limiting its use in research. In the SAP corpus~\cite{hasegawa2024community}, only a small fraction of each participant’s speech, i.e., roughly 30 read sentences, is rated by speech-language pathologists, compared with 350--400 read sentences and 50--80 spontaneous speech samples in total. This limited amount of labeled datasets for dysarthric speech may limit the performance of DSQA models, especially in terms of robustness. 

To fully leverage the SAP dataset, which contains a small labeled subset and a much larger unlabeled corpus, we propose a three-stage framework. Additionally, we propose to adopt the large-scale typical speech corpus LibriSpeech~\cite{panayotov2015librispeech} to enhance the speaker variability and acoustic environment of the training dataset and improve the robustness. In the first stage, a teacher regression model is trained on the limited labeled SAP subset using Whisper-large~\cite{radford2023robust} as the speech encoder, then used to generate pseudo-labels for the extensive unlabeled SAP samples. In the second stage, these pseudo-labeled samples are then combined with Librispeech and used for weakly-supervised pretraining. During this stage, we employ a label-aware contrastive learning approach inspired by~\cite{supcon} to better align the representations with perceptual quality labels. In the final stage, the pretrained representation layers are fine-tuned on the labeled SAP subset for the downstream regression task.

To demonstrate the effectiveness of our proposed methods, especially in terms of robustness to unseen test data, we build diverse test datasets with varying etiologies and languages, including UASpeech~\cite{kim08c_interspeech}, DysArinVox~\cite{zhang24l_interspeech}, EasyCall~\cite{turrisi21_interspeech}, EWA-DB~\cite{EWA-DB2023}, and NeuroVoz~\cite{mendes2024neurovoz}. Our baseline model, trained on the SAP labeled subset, achieves an utterance-level Spearman's Rank Correlation Coefficient (SRCC) of 0.719 on the SAP test set and an average speaker-level SRCC of 0.732 on the cross-domain test sets. Our proposed three-stage framework consistently improves over the baseline, reaching an average SRCC of 0.761 on the cross-domain test sets while preserving performance on the SAP dataset. We further demonstrate that weak supervision is essential for harmonizing the SAP and LibriSpeech datasets and improving cross-domain robustness. Our contributions are summarized as follows.
\begin{itemize}
    \item We propose a three-stage framework that gradually improves the labeling quality of the SAP dataset by using pseudo-labeling, followed by representation learning with a contrastive objective. This method fully leverages the large portion of unlabeled data in the SAP dataset.
    \item We propose to incorporate a large-scale typical speech dataset, LibriSpeech, into our DSQA model training pipeline, which improves the speaker and acoustic environment variability.
    \item We evaluate our method with various cross-domain dysarthric speech datasets with diverse labeling metrics and languages, demonstrating the effectiveness and robustness of our proposed methods.
    \item Through a rigorous ablation study, we demonstrate the importance of providing weak supervision during the representation learning stage and adopting the LibriSpeech dataset. 
    \item Model checkpoint and training code are publicly available: \url{https://github.com/JaesungBae/DA-DSQA}
\end{itemize}
\section{Related Works}

\subsection{Speech Quality Assessment}

Speech quality assessment (SQA) aims to predict perceived speech quality, typically represented by a mean opinion score (MOS). Traditional intrusive metrics such as PESQ~\cite{rix2001perceptual} and signal-based measures like SI-SNR~\cite{si-sdr} and STOI~\cite{stoi} require reference signals, and even codec-oriented intrusive metrics like WARP-Q~\cite{warpq} share this limitation, whereas modern generative applications (e.g., text-to-speech (TTS) and speech enhancement) increasingly rely on non-intrusive SQA (NI-SQA) methods. With the availability of large-scale MOS datasets and benchmarks, deep learning approaches have become the dominant paradigm for NI-SQA. Representative systems include DNSMOS~\cite{reddy2021dnsmos}, which predicts multi-dimensional MOS for the deep noise suppression (DNS) task, and UTMOS~\cite{saeki2022utmos}, which employs a strong-weak learner framework to model human perception of the naturalness of synthetic speech and achieved state-of-the-art performance in the VoiceMOS Challenge 2022~\cite{huang2022voicemos}. More recently, \cite{Halimeh2025OnTR} introduced a training-free NI-SQA metric based on neural codec latent representations. While DNSMOS and UTMOS have been applied to evaluate synthetic dysarthric speech~\cite{sanchez2025can, de2025objective}, they are primarily designed for healthy or synthetic speech, leaving their applicability to pathological speech unclear.
% \minje{BSS\_eval, SI-SNR, STOI, PEASS \url{https://dl.acm.org/doi/abs/10.1007/978-3-642-28551-6\_53}, WARP-Q, \url{https://scholar.google.com/citations?hl=en&user=Rs6wWLQAAAAJ&view_op=list_works&sortby=pubdate}, ``On the relation between speech quality and quantized latent representations of neural codecs"}

\subsection{Dysarthric Speech Quality Assessment}

Unlike conventional SQA, perceptual quality assessment of dysarthric speech often focuses on clinically relevant attributes such as intelligibility and other voice quality dimensions assessed by speech-language pathologists (SLPs). SpICE~\cite{venugopalan2023spice} automatically estimates human-perceived intelligibility of dysarthric speech using large-scale data from Project Euphonia~\cite{macdonald2021disordered}. 
More recently, \cite{narain2025voicequality} proposes voice quality probes trained on Speech Accessibility Project (SAP) data to predict multiple perceptual dimensions of dysarthric speech, including naturalness and intelligibility.

While these studies show that perceptual attributes of pathological speech can be learned from large-scale datasets using self-supervised representations, most approaches rely on foundation models~\cite{baevski2020wav2vec, hsu2021hubert, whisper} trained predominantly on healthy speech. As a result, the learned representations may be suboptimal for capturing dysarthric speech characteristics, highlighting the need for representations more sensitive to dysarthric speech.

\subsection{Contrastive Learning}
Contrastive learning is a powerful approach for representation learning that pulls similar samples together while pushing dissimilar ones apart. Self-supervised methods such as SimCLR~\cite{simclr} learn representations from unlabeled data by making augmented pairs of the same sample similar while treating others as negatives. This process enables the model to capture the intrinsic structure of the data. In speech, contrastive objectives underpin self-supervised models such as wav2vec 2.0~\cite{baevski2020wav2vec} and HuBERT~\cite{hsu2021hubert}, which show strong transferability across downstream tasks~\cite{Yang2021SUPERBSP, emobox} including DSQA~\cite{venugopalan2023spice, narain2025voicequality}.
However, purely self-supervised contrastive learning does not explicitly align representations with task-relevant perceptual attributes. Supervised contrastive learning (SupCon)~\cite{supcon} addresses this limitation by defining samples within the same label as positive pairs. \cite{Zha2022RankNContrastLC} further improves this approach for the regression task by contrasting samples based on the label order, and improves robustness, efficiency, and generalization. \cite{jsqa} proposes an SQA model with contrastive pretraining on audio pairs generated by injecting noise at perceptually similar SNR levels. However, applying such methods to DSQA is challenging due to highly limited and imbalanced labels. 

Motivated by these advances, we propose a weakly supervised contrastive learning strategy that leverages pseudo-label-informed similarity together with coarse pairing between typical and dysarthric speech. This enables task-aware representation learning while mitigating label noise from a label-constrained dysarthric speech dataset. 

\section{Background}
\subsection{Severity Level Prediction}
Severity-level prediction is a key dimension of dysarthric speech quality assessment (DSQA), aiding clinical assessment and providing auxiliary supervision for downstream tasks such as automatic dysarthric speech recognition (ASR)~\cite{severity_level_asr} and dysarthric speech generation with a TTS model~\cite {severity_level_tts}. However, collecting dysarthric speech data with severity-level annotations is costly as it requires expert clinical evaluation. Moreover, existing datasets suffer from labeling scale discrepancies and class imbalance, thus exhibiting inherently ambiguous decision boundaries between severity levels. 
To address these challenges, we formulate severity prediction as a regression task rather than classification, enabling soft estimation of speech impairment levels, following prior work~\cite{narain2025voicequality, Merler2025ClinicalAA}.

\begin{table}[t]
\centering
\caption{Summary of dataset information. Training is English-only, while cross-domain tests are multi-lingual. EN: English; ZH: Mandarin Chinese; IT: Italian; CS: Czech; SK: Slovak; ES: Spanish. ALS: Amyotrophic lateral sclerosis; CP: Cerebral palsy; DS: Down syndrome; PD: Parkinson’s disease; AD: Alzheimer's disease; MCI: Mild Cognitive Impairment. DysArinVox provides fine-grained annotations across 20 disorder categories (e.g., paralysis, leukoplakia, and polyps), and EasyCall includes dysarthria associated with PD, Huntington’s disease (HD), ALS, peripheral neuropathy, and myopathic or myasthenic lesions. LBL and Spk indicate the labeling method and the number of speakers, respectively. 
% Note that only about 10 hrs among 243 hrs of speech is labeled for the SAP train dataset. \todo{I think now the caption is too long.}
}
\vspace{\tablecaptionvspaceval}
\label{tab:dataset}
\tiny
\setlength{\tabcolsep}{4pt}
\renewcommand{\arraystretch}{1.08}
\footnotesize
% \begin{tabularx}{1.0\columnwidth}{@{}l l Y c r r@{}}
\resizebox{\linewidth}{!}{
\begin{tabular}{@{}l l l c r r@{}}
\toprule

\textbf{Lang.} & \textbf{Name} & \textbf{Etiology} & \textbf{LBL} & \textbf{Hours} & \textbf{\# Spk}\\
\midrule
% \hline
\rowcolor{gray!12}\multicolumn{6}{@{}l}{\textbf{Training Data}}\\
% \addlinespace[2pt]
% \midrule

\multirow{2}{*}{EN} & 
\multirow{2}{*}{SAP~\cite{hasegawa2024community} Train} &
\multirow{2}{*}{ALS, CP, DS, PD} &
\cmark & 10.8 & 318 \\
& &  & \xmark & 232.9 & 722 \\

EN & LibriSpeech~\cite{panayotov2015librispeech} & Healthy & -- & 921.7 & 2.3k \\

% \addlinespace[2pt]
% \midrule
\hline
\rowcolor{gray!12}\multicolumn{6}{@{}l}{\textbf{In-domain Evaluation Data}}\\
\addlinespace[2pt]
% \midrule
% \multicolumn{6}{@{}l}{\textit{In-domain}}\\
EN & SAP~\cite{hasegawa2024community} Test & \mbox{ALS, CP, DS, PD} & \cmark & 3.1 & 89 \\

% \addlinespace[2pt]
% \midrule
\hline
\rowcolor{gray!12}\multicolumn{6}{@{}l}{\textbf{Cross-domain Evaluation Data}}\\
\addlinespace[2pt]
% \midrule

% \multicolumn{6}{@{}l}{\textit{Out-of-domain (OOD)}}\\
EN    & UASpeech~\cite{kim08c_interspeech}        & CP           & Intellig. & 7.8  & 15  \\
ZH    & DysArinVox~\cite{zhang24l_interspeech}    & Mixed      & MOS       & 2.3  & 72  \\
IT    & EasyCall~\cite{turrisi21_interspeech}     & Mixed      & TOM       & 10.0 & 44  \\
CS/SK & EWA-DB~\cite{EWA-DB2023}                  & PD, AD, MCI  & MoCA      & 4.4  & 136 \\
ES    & NeuroVoz~\cite{mendes2024neurovoz}        & PD           & H-Y       & 1.7  & 98  \\

\bottomrule
\end{tabular}

}
\vspace{-0.1in}
% \vspace{3pt}
\end{table}

% \subsection{Datasets}
Table~\ref{tab:dataset} shows the summary of the training and test datasets used in this work. For training, we use the SAP dataset~\cite{hasegawa2024community} (covering data collected and annotated through January 31, 2025). Although SAP is one of the largest available dysarthric speech corpora, human expert-annotated samples, based on clinically or perceptually relevant metrics, remain limited, as noted in Section~\ref{sec:intro}. 
Out of the dataset's 32 perceptual rating dimensions, we select \emph{naturalness} and \emph{intelligibility} as the primary indicators of severity. Each is rated on a 7-point scale, where 1 denotes typical speech, and 7 indicates severe dysarthria. We define the target severity score as the average of these two ratings. Considering only utterances shorter than 15 seconds, 10.8 hours of speech are labeled, which is only about 4.6\% of the 232.9 hours of unlabeled speech. 
% ; the effectiveness of this design choice is examined in Section~\ref{sec:ablation}.
% \minje{Introduce the other datasets used for testing.}
% The summary of the training and test datasets is shown in Table~\ref{tab:dataset}. 
The training splits of the LibriSpeech~\cite{panayotov2015librispeech} dataset are used as an additional normal speech dataset for contrastive learning.

In addition to the SAP in-domain test, we further evaluate generalizability using cross-domain datasets: UASpeech~\cite{kim08c_interspeech}, DysArinVox~\cite{zhang24l_interspeech}, EasyCall~\cite{turrisi21_interspeech}, EWA-DB~\cite{EWA-DB2023}, and NeuroVoz~\cite{mendes2024neurovoz}. These dysarthric speech corpora span different languages and etiologies, and adopt labeling schemes that differ substantially from SAP, making cross-dataset severity prediction extremely challenging. Specifically, UASpeech provides overall speech intelligibility scores derived from word-recognition accuracy by five native listeners; DysArinVox reports MOS ratings based on subjective perceptual evaluations of Mandarin dysphonic and dysarthric speech; EasyCall provides therapy outcome measure (TOM) scores assigned by experienced neurologists to categorize the clinical severity levels; EWA-DB provides Montreal cognitive assessment (MoCA) scores reflecting cognitive health status; NeuroVoz provides Hoehn and Yahr (H-Y) scale stages to categorize the progressive severity of Parkinson’s disease symptoms and their impact on motor function. 

As these cross-domain datasets are labeled per speaker rather than per utterance, we predict utterance-level severity scores and average them per speaker. For EasyCall, we merge the official train and test splits to create a larger evaluation set with more speakers. For EWA-DB, we restrict evaluation to speakers with AD, PD, or MCI. Hyperparameters are selected solely based on the validation splits of SAP and EasyCall.

\subsection{Contrastive Learning}

\begin{figure*}[h!]
  \centering
  % \vspace{-0.35in}
  \includegraphics[width=1.0\linewidth]{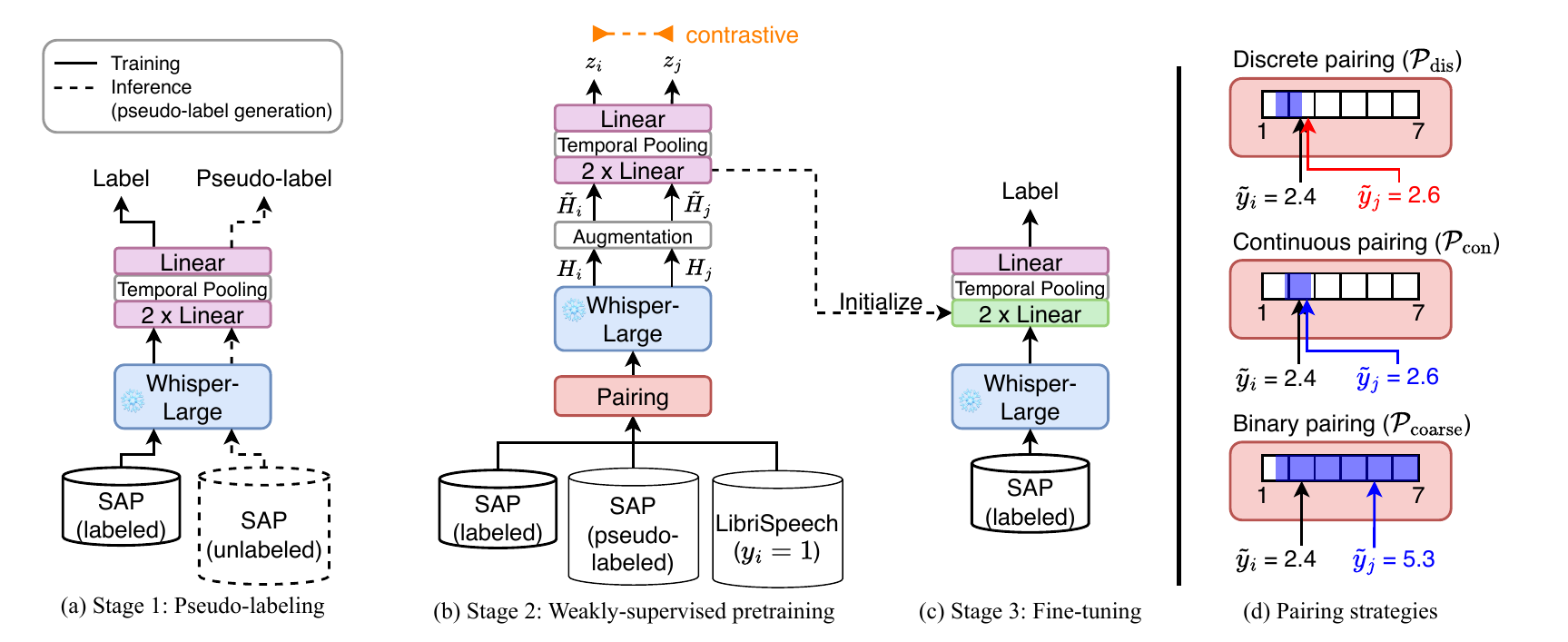}
  % \vspace{-0.07in}
  \caption{
  (a--c) Illustration of the three-stage framework with weakly supervised pretraining, and (d) the proposed pairing strategies for weakly supervised contrastive learning. 
(a) \textbf{Stage 1:} A regression model is trained on the labeled SAP dataset (3\% of the total), and pseudo-labels are generated for the unlabeled portion. 
(b) \textbf{Stage 2:} Three linear layers are trained with weakly supervised contrastive losses. Pseudo-labeled SAP data from Stage~1 and LibriSpeech are additionally used, with LibriSpeech assigned a label of one (healthy). 
(c) \textbf{Stage 3:} The first two linear layers from Stage~2 are fine-tuned with an additional final linear layer using the labeled SAP dataset. 
(d) In \textbf{Stage 2}, positive pairs for contrastive learning are generated in three ways: discretizing the label, thresholding the distance between continuous labels, or dividing the label range into a binary dichotomy. With anchor $\tilde{y}_i=2.4$, samples with $\tilde{y}_j$ in the blue region form positive pairs: for example, $\tilde{y}_j=2.6$ is negative under discrete pairing but positive under continuous pairing. Binary pairing divides the data into two groups, providing weaker supervision.
  % (a--c) Illustration of the three-stage framework with weakly supervised pretraining, and (d) our proposed pairing methods for weakly supervised contrastive learning. (a) \textbf{Stage 1:} We train a regression model using the labeled SAP dataset (3\% of the total) and then generate pseudo-labels for the unlabeled portion. (b) \textbf{Stage 2:} We train three linear layers with weakly supervised contrastive losses. The pseudo-labeled SAP data from Stage 1 and LibriSpeech are additionally used, with LibriSpeech assigned a label of one (healthy). (c) \textbf{Stage 3:} We take the first two linear layers from Stage 2 and fine-tune them, together with an additional final linear layer, using the labeled SAP dataset. (d) In \textbf{Stage 2}, we generate positive pairs 
  % for contrastive learning in three different ways: by discretizing the label,
  % by thresholding the distance between continuous labels, or by 
  % categorizing the complete label range into a binary dichotomy. 
  % With anchor $\tilde{y}_i=2.4$, the data with $\tilde{y}_j$ in the blue area becomes the positive pairs. $\tilde{y}_j=2.6$ becomes negative and positive pairs in the discrete and continuous pairing strategy, respectively. With binary pairing, we roughly divide the data into two groups, giving weaker supervision to the model.
}
\vspace{-0.1in}
  \label{fig:model}
\end{figure*}

Our approach builds upon the SimCLR~\cite{simclr} contrastive learning frameworks, and the extended version of it with label supervision, SupCon~\cite{supcon}. Both methods generate two augmented views of each sample within a mini-batch, resulting in two copies of the batch. In our framework, to reduce the computational burden of performing inference on the Whisper-large model at each iteration, the augmentation process is defined in the feature vectors extracted by the Whisper encoder. Let the extracted feature vector be $H_i=\text{Whisper}(\mathbf{x}_i)$. For a randomly sampled mini-batch of size $B$, we obtain $\{H_i, y_i\}_{i=1}^{B}$, where $H_i$ is normalized. Note that $y_i$ may correspond to the pseudo-label $\bar{y}_i$ for pseudo-labeled data; for simplicity, we denote both as $y_i$. We construct an augmented batch $\{\tilde{H}_j, \tilde{y}_j\}_{j=1}^{2B}$, where $\tilde{H}_i$ and $\tilde{H}_{B+i}$ are two independently augmented versions of $H_i$ for $i=1,\dots,B$, and $\tilde{y}_i=\tilde{y}_{B+i}=y_i$. The final representation $\mathbf{z}_i$ is then obtained by passing $\tilde{H}_i$ through two linear layers, followed by statistical temporal pooling and a final linear layer (Figure~\ref{fig:model}).

% \noindent\textbf{Self-supervised contrastive learning:} \todo{This section might be moved to a background section.} \minje{I think it's also okay to keep it here for a more in-context explanation of SimCLR. A bit too heavy to talk about this in the lit review section, I suppose. (re: okay, I'll keepe it here Thanks!)} 
SimCLR~\cite{simclr} is a self-supervised method that treats two different augmented versions of the same representation as a positive pair, while all other samples in the batch serve as negative pairs. Let $i \in I \coloneqq \{1, \dots, 2B\}$ denote the index of a sample in the augmented batch. For a given index $i$, there is always a positive pair for it within the same batch, whose index $k(i)$ is defined as the index of the other augmented sample originating from the same source sample.
% \begin{equation}
% k(i) =
% \begin{cases}
% i + B, & \text{if } 1 \le i \le B, \\
% i - B, & \text{if } B < i \le 2B.
% \end{cases}
% \end{equation}
The normalized temperature-scaled cross-entropy (NT-Xent) loss is defined as
\begin{equation}
\mathcal{L}_{\text{SimCLR}}
=
-\frac{1}{2B}
\sum_{i \in I}
\log
\frac{
\exp\left(\mathbf{z}_i^\top \mathbf{z}_{k(i)} / \tau\right)
}{
\sum\limits_{j \in A(i)}
\exp\left(\mathbf{z}_i^\top \mathbf{z}_j / \tau\right)
}
\end{equation}
where $\tau > 0$ is a temperature parameter and $A(i) \coloneqq I \setminus \{ i \}$.

However, $\mathcal{L}_{\text{SimCLR}}$ does not utilize label information $y_i$, although it could be informative for the downstream task. Instead, relies solely on the inherent data structure and cannot explicitly guide the representation space toward task-relevant features. To this end, SupCon proposes a method to consider the data samples with the same label as positive pairs, rather than considering all the other data samples, but the augmented version of it as negative. The positive pair set for index $i$ is defined as follows:
\begin{equation}
\mathcal{P}_{\text{sup}}(i)
=
\left\{
j \in A(i) \mid
\tilde{y}_j=\tilde{y}_i\right\}.
\end{equation}
Using these positive pairs, SupCon optimizes the model with the following modified objective based on $\mathcal{L}_{\text{SimCLR}}$:
\begin{equation}\label{eq:l_fine}
\mathcal{L}_{\text{sup}}=-\frac{1}{2B}\sum_{i \in I}\frac{1}{|\mathcal{P}_{\text{sup}}(i)|}\sum_{p \in \mathcal{P}_{\text{sup}}(i)}\log\frac{\exp\left(\mathbf{z}_i^\top \mathbf{z}_{p} / \tau\right)}{\sum\limits_{j \in A(i)}\exp\left(\mathbf{z}_i^\top \mathbf{z}_j / \tau\right)}.
\end{equation}

% \begin{wrapfigure}[17]{r}{0.4\textwidth}

\section{Proposed Method}
Our goal is to develop a robust and generalizable DSQA model. Due to the scarcity of labeled dysarthric speech data, our key motivation is to leverage large amounts of unlabeled dysarthric speech alongside large-scale typical speech. This allows the model to be exposed to diverse speaker identities and acoustic environments. However, effectively utilizing such unlabeled data is challenging because the absence of reliable severity annotations prevents direct supervised optimization. We propose a three-stage framework: (1) pseudo-label generation using a supervised regression model trained on labeled data, (2) weakly supervised representation learning via contrastive objectives, and (3) fine-tuning a regression model on labeled data.

\begin{figure}[t]
     \centering
     \begin{subfigure}[b]{0.49\linewidth}
         \centering
         \includegraphics[width=\textwidth]{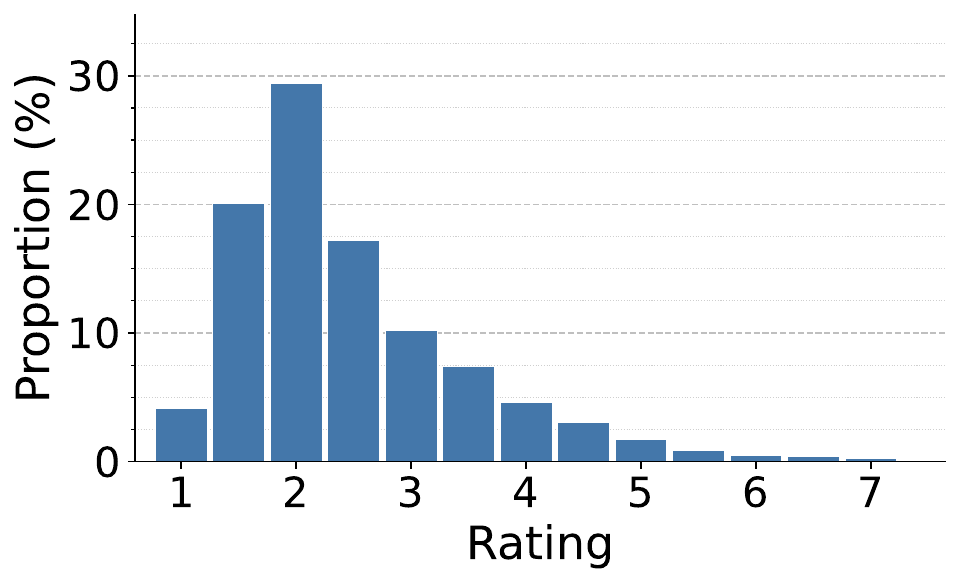}
         \vspace{-0.05in}
         % \caption{$\mathcal{D}_\text{labeled}$}
         \label{fig:histogram:orig}
     \end{subfigure}
     % \qquad
     \begin{subfigure}[b]{0.49\linewidth}
         \centering
         \includegraphics[width=\textwidth]{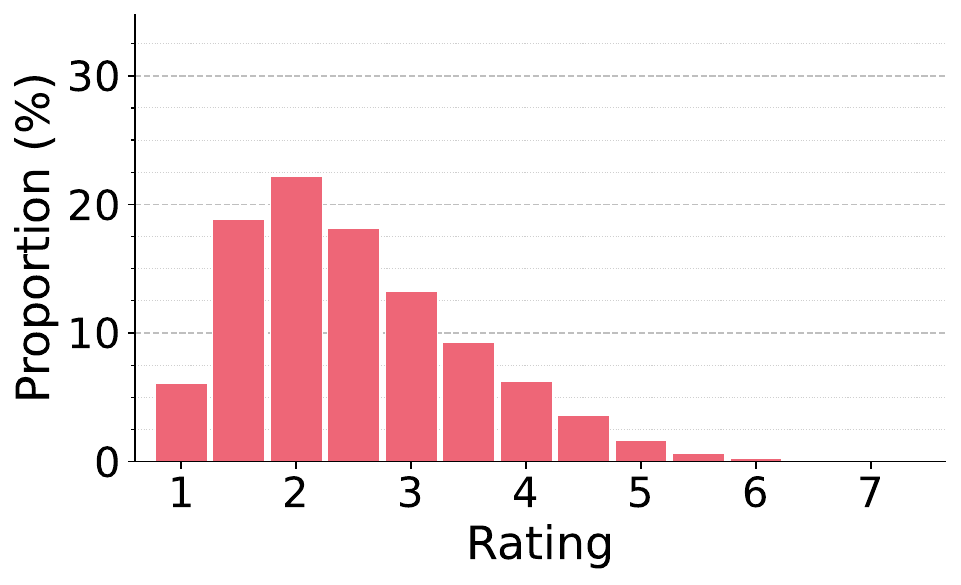}
         \vspace{-0.05in}
         % \caption{$\mathcal{D}_\text{pseudo}$}
         \label{fig:histogram:pseudo}
     \end{subfigure}
        \vspace{-0.3cm}
        \caption{Histogram of label proportions in $\mathcal{D}_\text{labeled}$ (left) and $\mathcal{D}_\text{pseudo}$ (right).}
        \vspace{\vspaceval}
        \label{fig:histogram}
\end{figure}
\subsection{Stage 1: Pseudo-Label Generation}
We first train a regression model using the labeled subset of SAP. Let $\mathbf{x}$ denote an input speech utterance and $y \in [1,7]$ its corresponding severity label. Denote the labeled and unlabeled datasets as
\[
\mathcal{D}_{\text{labeled}} = \{(\mathbf{x}_i, y_i)\}_{i=1}^{N}, \quad
\mathcal{D}_{\text{unlabeled}} = \{\mathbf{x}_i\}_{i=1}^{M},
\]
where $M \gg N$. Our regression model architecture is shown in Figure~\ref{fig:model}. We use the encoder portion of Whisper-large as a speech foundation model to extract feature representations from the input waveform, which is frozen during training. The extracted frame-level features are passed to the trainable component: two linear layers followed by statistical temporal average pooling~\cite{ecapa-tdnn}. The pooled representation is then mapped to a scalar severity score via a final linear layer. After training the regression model on $\mathcal{D}_{\text{labeled}}$, we use it to predict severity scores for $\mathcal{D}_{\text{unlabeled}}$, producing pseudo-labels $\bar{y}_i$. This yields a pseudo-labeled dataset:
\[
\mathcal{D}_{\text{pseudo}} = \{(\mathbf{x}_i, \bar{y}_i)\}_{i=1}^{M}.
\]

The proportion histogram of $\mathcal{D}_{\text{labeled}}$ and the resulting $\mathcal{D}_{\text{pseudo}}$ over $y$ is shown in Figure~\ref{fig:histogram}. The initial regression model trained on $\mathcal{D}_\text{labeled}$ achieves an SRCC of 0.719 on the SAP test set (Table~\ref{tab:results-main}), indicating reasonably reliable predictions. However, these pseudo-labels are inherently imperfect. Training another regression model on $\mathcal{D}_{\text{pseudo}}$ is ineffective because it would largely replicate the data distribution of $\mathcal{D}_{\text{labeled}}$, while inheriting the bias from the initial model instead of fully leveraging the diversity in $\mathcal{D}_{\text{unlabeled}}$. Hence, we propose to use pseudo-labels only to construct weak supervision signals for representation learning in Stage 2.

\subsection{Stage 2: Weakly Supervised Representation Learning}
The main objective of this stage is to learn a more structured representation space that is both task-relevant and generalizable for robust assessment of dysarthric speech quality.
Contrastive learning has been widely adopted for representation learning~\cite{simclr, supcon, Zha2022RankNContrastLC} and has been shown to improve downstream generalization. Inspired by this, we employ contrastive learning to adapt the Whisper encoder’s representation space to our target task using $\mathcal{D}_{\text{labeled}}$ and $\mathcal{D}_{\text{pseudo}}$. Although SAP is relatively large, it consists solely of dysarthric speech, which may limit variability in speaker identities and acoustic environments. To further increase diversity, we additionally incorporate LibriSpeech~\cite{panayotov2015librispeech}, as a large-scale typical speech dataset. Since a label of 1 in SAP indicates speech with no noticeable impairment, we assign the typical speech samples this label:
\[
\mathcal{D}_{\text{Libri}}=\{(\mathbf{x}_i, 1)\}_{i=1}^{K}
\]
where $K$ denotes the number of typical samples, we analyze the impact of incorporating LibriSpeech in Section~\ref{sec:ablation:data_loss}.

\subsubsection{Proposed weakly supervised contrastive learning}

In our pilot study, we observe that the simply applying SimCLR with our mixed datasets $\mathcal{D}_\text{labeled}$, $\mathcal{D}_\text{pseudo}$, and $\mathcal{D}_\text{Libri}$ was unable to form a meaningful representation space for the downstream task. This is likely because of the different data characteristics between the SAP and LibriSpeech datasets. Applying the SupCon directly is also problematic because of the continuous nature of the predicted score of the regression model for $\mathcal{D}_\text{pseudo}$. 

% \noindent\textbf{Fine-grained supervision in contrastive learning:} Inspired by~\cite{}, we propose a contrastive learning method that incorporates label (or pseudo-label) information to inject downstream task-specific supervision into the representation space. 
To this end, we first propose a supervised contrastive learning method to discretize the label $y_i$ to the nearest integer and define positive pairs among samples sharing the same discretized label:
\begin{equation}
\left\lfloor y \right\rfloor
=
\max \{ n \in \mathbb{Z} \mid n \le y \}.
\end{equation}
\begin{equation}
\mathcal{P}_{\text{dis}}(i)
=
\left\{
j \in A(i) \mid
\left\lfloor y_j + \tfrac{1}{2} \right\rfloor
=
\left\lfloor y_i + \tfrac{1}{2} \right\rfloor
\right\}.
\end{equation}
% following modified objective based on $\mathcal{L}_{\text{SimCLR}}$:
% \begin{equation}
% \mathcal{L}_{\text{dis}}
% =
% -\frac{1}{2B}
% \sum_{i \in I}
% \frac{1}{|\mathcal{P}_{\text{dis}}(i)|}
% \sum_{p \in \mathcal{P}_{\text{dis}}(i)}
% \log
% \frac{
% \exp\left(\mathbf{z}_i^\top \mathbf{z}_{p} / \tau\right)
% }{
% \sum\limits_{j \in A(i)}
% \exp\left(\mathbf{z}_i^\top \mathbf{z}_j / \tau\right)
% }.
% \label{eq:l_fine}
% \end{equation}

However, this discretization introduces discontinuities near label boundaries. For example, labels 2.4 and 2.6 are expected to be more similar than 2.4 and 1.7, yet the former pair is not considered positive while the latter is. To address this, we propose an alternative pairing strategy based on label distance. Specifically, we define positive pairs as those whose label difference is smaller than a threshold $\alpha$:
\begin{equation}
\mathcal{P}_{\text{con}}(i)
=
\left\{
j \in A(i) \mid
|y_j - y_i| < \alpha
\right\}.
\end{equation}

However, pseudo-labels may be inaccurate, and excessive reliance on them can introduce misleading supervision. Additionally, as shown in Figure~\ref{fig:histogram}, the distribution of labels $y$ and $\bar{y}$ are skewed toward lower rating values, meaning that constructing positive pairs from $\mathcal{P}_\text{dis}$ and $\mathcal{P}_\text{con}$ risks overfitting to low-score samples. To mitigate these issues, we introduce a coarser supervisory signal by grouping samples into two categories: dysarthric speech and typical speech. Given a threshold $\beta$, a sample is considered dysarthric if $y_i > \beta$, and typical otherwise. Based on this binary grouping, the positive pair set is defined as
\begin{equation}\label{eq:corse}
\mathcal{P}_{\text{coarse}}(i)
=
\left\{
j \in A(i) \mid
% s(y_j;\beta)=s(y_i;\beta)
\mathbf{1}\{y_j > \beta\} = \mathbf{1}\{y_i > \beta\}
\right\},
\end{equation}
where $\mathbf{1}\{\cdot\}$ denotes the indicator function, which equals 1 if the condition inside the braces is true and 0 otherwise. By proposing this coarse binary grouping strategy, we can expect the low severity level samples to act as a bridge between the LibriSpeech and SAP dataset, while high severity level samples will also learn the structured representation space, including the task-relevant information.

For each positive pair, we compute the corresponding loss objectives $\mathcal{L}_{\text{dis}}$, $\mathcal{L}_{\text{con}}$, and $\mathcal{L}_{\text{coarse}}$ by replacing $\mathcal{P}_\text{sup}$ with $\mathcal{P}_\text{dis}$, $\mathcal{P}_\text{con}$, and $\mathcal{P}_\text{coarse}$, respectively. The thresholds $\alpha$ and $\beta$ are set to 0.5 and 1.5.\\[3pt]

\noindent\textbf{Temperature value $\tau$:} 
The temperature value $\tau$ is an important factor that controls the strength of the supervision of the contrastive learning as suggested in the literature~\cite{supcon, Wang2020UnderstandingTB}. If $\tau$ is small, the loss function encourages hard decisions even on confusing examples, while big $\tau$ values allow for such confusions, leading to a more gradually distributed feature space between different severity levels. As we incorporate a new dataset $\mathcal{D}_\text{Libri}$, a small $\tau$ can lead to a trivial separation between the features of $\mathcal{D}_\text{Libri}$ and those of the SAP dataset $\mathcal{D}_\text{labeled}$ and $\mathcal{D}_\text{unlabeled}$. We investigate different choices of $\tau$ and observe that larger values tend to provide more robust feature spaces (Section~\ref{sec:ablation:tau}). 

\subsubsection{Variance Regularization}
In addition to the contrastive objective, we incorporate the variance regularization term from VICReg~\cite{vicreg} to prevent representational collapse. This regularizer penalizes embedding dimensions whose batch-wise standard deviation falls below a predefined threshold $\gamma$:

\begin{equation}
\mathcal{L}_{\text{var}} =
\frac{1}{d} \sum_{k=1}^{d}
\max \left(0, \gamma - \sqrt{\operatorname{Var}(\mathbf{z}_{:,k}) + \varepsilon} \right).
\label{eq:vicreg_var}
\end{equation}

where $\mathbf{z} \in \mathbb{R}^{2B \times d}$ denotes the embedding matrix for a batch of $2B$ samples with embedding dimension $d$, and $\mathbf{z}_{:,k}$ represents the $k$-th feature dimension across the batch. This hinge-style penalty encourages informative and well-dispersed representations across feature dimensions. We set $\gamma=1.0$.

\subsubsection{Final Objective}
The overall representation learning objective in Stage 2 is defined as
\begin{equation}
\mathcal{L}_{\text{Stage2}} =
\mathcal{L}_{\text{contrast}} + \lambda \mathcal{L}_{\text{var}},
\end{equation}
where $\lambda$ controls the strength of the variance regularization, and $\mathcal{L}_{\text{contrast}}$ can be one of $\mathcal{L}_{\text{dis}}$, $\mathcal{L}_{\text{con}}$, or $\mathcal{L}_{\text{coarse}}$. We set $\lambda = 0.1$.

\subsection{Stage 3: Fine-Tuning for Regression}

In the final stage, we initialize the first two adaptor layers using the pretrained weights obtained from Stage 2. A randomly initialized linear layer is added on top to predict continuous severity scores. The entire model is then fine-tuned on $\mathcal{D}_{\text{labeled}}$. By decoupling representation learning from regression optimization, our framework leverages large-scale unlabeled data while mitigating the adverse effects of pseudo-label noise.

\section{Experiments}

\subsection{Experimental setup}
Whisper-large-v3~\cite{whisper} is adopted as the backbone for feature extraction and is frozen throughout training. Whisper features are extracted after applying voice activity detection (VAD)~\cite{silero-vad} to the original speech signals following~\cite{das2025improved}.

For Stage~1, two linear layers are applied, followed by statistical temporal pooling~\cite{ecapa-tdnn}, which aggregates frame-level representations into an utterance-level representation. A final linear layer maps the representation to a severity score. ReLU activation and dropout ($p=0.1$) are applied after each linear layer. The first two layers have output dimensions of 320. Training uses Huber loss ($\delta=0.5$) with a batch size of 32 and a learning rate of $10^{-4}$, optimized with AdamW~\cite{adamw}. To mitigate class imbalance, we apply label-weighted random sampling. The model is trained for 10 epochs, selecting the best checkpoint based on SAP validation SRCC.

% Whisper-large-v3 model is adopted as the backbone for feature extraction, which is frozen throughout the whole process. The Whisper feature is extracted after performing the voice activity detection (VAD)~\cite{silero-vad} of the original speech signals following~\cite{dysarthric-speech-enhancement}. 

% For stage 1, two linear layers are applied afterward, followed by statistical temporal pooling~\cite{ecapa-tdnn}, which aggregates the temporal representations into an utterance-level representation. Finally, one linear layer converts the output into a severity score. ReLU activation and dropout with a probability of 0.1 are applied to every linear layer. The output dimensions of the first two layers are both 320. Training uses Huber loss ($\delta=0.5$) with a batch size of 32 and a learning rate of $10^{-4}$, optimized with AdamW~\cite{adamw}. To mitigate class imbalance in the rating distribution, we apply label-weighted random sampling. We train for 10 epochs, selecting the best checkpoint by SAP validation set SRCC. 

For stage 2, to generate an augmented view, two augmented views are generated by composing Gaussian noise ($\sigma=0.01$) with stochastic time masking (up to 20\% of frames) and random temporal cropping ($\ge$70\% of the sequence), each applied with 50\% probability to the Whisper feature. Two linear layers with dimension 320, temporal pooling, and one linear layer with 128 output dimension are applied to convert $\tilde{H}$ into $\mathbf{z}$. Training uses Adam with learning rate $10^{-3}$ and weight decay $10^{-5}$, and we train for 2 epochs. We observe that as training progresses, performance generally drops. We assume that heavy fine-tuning of the Whisper features—which are already rich and well-structured—may harm their effectiveness. For the contrastive learning methods, we conduct experiments with $\tau \in \{0.1, 1.0, 10.0, 50.0, 100.0\}$ and select the optimal value based on the average SRCC score on the validation sets after Stage 3. For proposed models using $\mathcal{L}_\text{dis}$, $\mathcal{L}_\text{con}$, and $\mathcal{L}_\text{coarse}$, the selected $\tau$ values were 1.0, 0.1, 0.1, and 10.0, respectively.

All Stage~3 settings are identical to those in Stage~1. Experiments are repeated with five random seeds for stability.

\subsubsection{Evaluation metrics}
To evaluate the monotonic and linear consistency of our regression models, we adopt two correlation-based metrics: SRCC and Pearson Correlation Coefficient (PCC). SRCC assesses whether the predicted scores preserve the relative ordering of samples, while PCC quantifies the strength of their linear relationship. Unlike absolute error metrics (e.g., MAE), both SRCC and PCC are invariant to affine transformations (scaling and shifting), making them particularly suitable for cross-domain test cases where label ranges may differ.
% We exclude direct error calculation metrics such as MAE as they do not generalize to the different label scales in the out-of-domain test cases.

\subsubsection{Comparison models}
\texttt{DNSMOS}~\cite{reddy2021dnsmos} employs a CNN-based multi-stage self-teaching framework to predict non-intrusive MOS. Three SSL-based comparison models are also considered. \texttt{UTMOS}~\cite{saeki2022utmos} fine-tunes a wav2vec 2.0~\cite{baevski2020wav2vec} encoder with a BLSTM and linear layers for frame-level prediction, with utterance-level scores obtained by averaging. The linear classifier in \texttt{SpICE}~\cite{venugopalan2023spice} is built on 12th-layer (768-d) representations from a wav2vec 2.0 backbone, with final scores computed as a label-weighted average of predicted class probabilities~\cite{narain2025voicequality}. The \texttt{HuBERT Probe}~\cite{narain2025voicequality} is reproduced by training a LASSO regression probe on a HuBERT-large (1024-d) backbone using $\mathcal{D}_\text{labeled}$, with VAD-based silence trimming and the same class-weighted sampler as our method. The LASSO $\ell_1$ coefficient is set to $1\mathrm{e}{-7}$, selected by in-domain validation over $\{0, 1\mathrm{e}{-7}, 1\mathrm{e}{-6}, 1\mathrm{e}{-5}\}$.
% footnote{https://huggingface.co/facebook/hubert-large-ll60k}

Within our three-stage framework, we define three additional comparison models: \texttt{Baseline}, \texttt{SimCLR}, and Rank-N-Contrast (\texttt{RNC})~\cite{Zha2022RankNContrastLC}. \texttt{Baseline} is a regression model fine-tuned on Whisper-large encoder features without pseudo-labeling or contrastive learning. \texttt{SimCLR} applies contrastive learning with $\mathcal{L}_\text{SimCLR}$ and does not require pseudo-labels. \texttt{RNC} uses a regression-oriented contrastive objective that contrasts samples based on their ranking in the label space to learn continuous, order-preserving representations. For \texttt{SimCLR} and \texttt{RNC}, the temperature $\tau$ is selected on the validation set and set to 0.1 and 100.0, respectively.

\begin{table*}[t]
\caption{Performance comparison on in-domain and cross-domain test sets. ``Average'' in cross-domain indicates the average performance across all cross-domain datasets. Bold and underlined values denote the best and second-best results. Across five seeds, the standard deviations of the in-domain and average cross-domain SRCCs for Proposed ($\mathcal{L}_{\text{coarse}}$) are 0.0021 and 0.0041, respectively.}
\vspace{\tablecaptionvspaceval}
\centering
\small
\resizebox{\textwidth}{!}{
% \scriptsize
\renewcommand{\arraystretch}{1.2}
\begin{tabular}{l|ccc|cc|cc|cc|cc|cc|cc|cc}
\toprule
\multirow{3}{*}{\textbf{Model}} & \multicolumn{3}{c|}{\multirow{2}{*}{\textbf{Stage}}} & \multicolumn{2}{c|}{\textbf{In-domain}} 
& \multicolumn{12}{c}{\textbf{Cross-domain}} \\
% \cmidrule(lr){2-7} \cmidrule(lr){8-19}
\cline{5-18}
~ & ~ & ~ & ~ & \multicolumn{2}{c|}{SAP} 
& \multicolumn{2}{c|}{UASpeech}
& \multicolumn{2}{c|}{DysArinVox}
& \multicolumn{2}{c|}{EasyCall}
& \multicolumn{2}{c|}{EWA-DB}
& \multicolumn{2}{c|}{NeuroVoz}
& \multicolumn{2}{c}{Average}
\\
\cline{2-18}
& 1 & 2 & 3
& SRCC & PCC
& SRCC & PCC
& SRCC & PCC
& SRCC & PCC
& SRCC & PCC
& SRCC & PCC
& SRCC & PCC\\
\hline
DNSMOS~\cite{reddy2021dnsmos} & ~ & ~ & ~ & 0.186 & 0.278 & 0.750 & 0.742 & 0.370 & 0.502 & 0.041 & 0.061 & -0.274 & -0.294 & -0.361 & -0.363 & 0.105 & 0.130\\
UTMOS~\cite{saeki2022utmos} & ~ & ~ & ~ & 0.489 & 0.499 & 0.962 & 0.912 & 0.521 & 0.525 & -0.051 & -0.108 & 0.328 & 0.345 & -0.028 & -0.065 & 0.346 & 0.322 \\
SpICE~\cite{venugopalan2023spice} & ~ & ~ & ~ & 0.473 & 0.505 & 0.936 & 0.926 & 0.538 & 0.463 & 0.205 & 0.237 & 0.393 & 0.366 & 0.180 & 0.201 & 0.450 & 0.439\\
HuBERT Probe~\cite{narain2025voicequality} & ~ & ~ & ~ & 0.531 & 0.620 & 0.927 & 0.924 & 0.279 & 0.334 & 0.604 & 0.704 & 0.588 & 0.487 & \textbf{0.705} & \textbf{0.706} & 0.621 & 0.631\\

\hline
Baseline & ~ & ~ & \cmark & 0.719 & \underline{0.755} & 0.949 & 0.963 & 0.578 & 0.619 & 0.849 & 0.783 & 0.709 & 0.686 & 0.575 & 0.601 & 0.732 & 0.730 \\
% Baseline + pseudo-label & \cmark & ~ & \cmark & 0.719 & 0.755 & 0.949 & 0.963 & 0.578 & 0.619 & 0.849 & 0.783 & 0.709 & 0.686 & 0.575 & 0.601 & 0.732 & 0.73\\
SimCLR & ~ & \cmark & \cmark & 0.716 & 0.739 & 0.951 & 0.960 & \underline{0.628} & \textbf{0.635} & \textbf{0.902} & \textbf{0.852} & 0.663 & 0.654 & 0.576 & 0.595 & \underline{0.744} & \underline{0.739} \\
Rank-N-Contrast & \cmark & \cmark & \cmark & \textbf{0.726} & \textbf{0.758} & \underline{0.959} & \underline{0.972} & 0.564 & 0.601 & 0.868 & 0.803 & \textbf{0.714} & \underline{0.693} & 0.577 & 0.592 & 0.736 & 0.732\\

\hline
Proposed ($\mathcal{L}_{\text{dis}}$)& \cmark & \cmark & \cmark & 
\underline{0.724} & 0.754 & 0.948 & 0.962 & 0.583 & 0.626 & 0.838 & 0.78 & 0.698 & 0.673 & 0.572 & 0.602 & 0.728 & 0.729\\
Proposed ($\mathcal{L}_{\text{cont}}$)& \cmark & \cmark & \cmark & 0.722 & \underline{0.755} & 0.935 & 0.956 & 0.558 & 0.599 & \underline{0.874} & \underline{0.811} & 0.677 & 0.659 & 0.516 & 0.553 & 0.712 & 0.716\\
Proposed ($\mathcal{L}_{\text{coarse}}$)& \cmark & \cmark & \cmark & 0.716 & 0.750 & \textbf{0.975} & \textbf{0.976} & \textbf{0.631} & \underline{0.632} & 0.872 & 0.810 & \underline{0.711} & \textbf{0.695} & \underline{0.617} & \underline{0.630} & \textbf{0.761} & \textbf{0.749} \\
% \hline
% w/o pseudo-labeling & ~ & \cmark & \cmark & 0.681 & 0.74 & 0.96 & 0.968 & 0.604 & 0.654 & 0.902 & 0.85 & 0.67 & 0.667 & 0.596 & 0.621 & 0.746 & 0.752 \\
% w/o $\mathcal{D}_{\text{Libri}}$ & \cmark & \cmark & \cmark &  0.653 & 0.724 & 0.939 & 0.971 & 0.588 & 0.648 & 0.859 & 0.837 & 0.701 & 0.686 & 0.608 & 0.637 & 0.739 & 0.756\\
% w/o $\mathcal{L}_{\text{var}}$ & \cmark & \cmark & \cmark & 0.698 & 0.751 & 0.967 & 0.975 & 0.628 & 0.655 & 0.883 & 0.837 & 0.702 & 0.681 & 0.557 & 0.586 & 0.747 & 0.747\\
\bottomrule
\end{tabular}
}
% \vspace{\vspaceval}
\label{tab:results-main}
\end{table*}

\subsection{Results}
\label{sec:result}
The test results on the in-domain SAP dataset and various multilingual cross-domain datasets are shown in Table~\ref{tab:results-main}. Overall, our \texttt{Baseline} model consistently outperforms existing NI-SQA (\texttt{DNSMOS}, \texttt{UTMOS}) and DSQA (\texttt{SpICE}, \texttt{HuBERT Probe}) models on all datasets except NeuroVoz, demonstrating the effectiveness of adapting the estimator to dysarthric severity levels and the robustness of the Whisper-large encoder. Although \texttt{DNSMOS} and \texttt{UTMOS} perform well in conventional SQA~\cite{sanchez2025can, de2025objective}, they generalize poorly to dysarthric speech severity prediction. In particular, \texttt{DNSMOS} exhibits low correlation with human-rated severity levels (avg SRCC$<$0.2). \texttt{UTMOS} shows relatively high correlation on MOS-style labeled datasets (e.g., DysArinVox) but drops significantly on others. Among existing DSQA metrics, \texttt{SpICE} is originally designed as a classification model, and converting it into a regression model may introduce result leakage. This highlights the limitations and challenges of adapting it to various datasets. \texttt{HuBERT Probe} achieves the best performance on NeuroVoz, possibly due to overfitting to PD speech, but shows a clear gap compared to \texttt{Baseline} on most other datasets.

Compared to the \texttt{Baseline}, \texttt{SimCLR} performs worse on the in-domain test but generally performs better on cross-domain tests. This is expected: since the \texttt{Baseline} is only exposed to the SAP dataset during training, it may be overfitted to that dataset. While it achieves reasonable performance on the cross-domain datasets, likely due to the generally high quality of the SAP data, its performance remains suboptimal. By exposing the model to unlabeled SAP and LibriSpeech datasets in Stage 2, we introduce more than 11.5K speakers and 1.1K hours of additional data, thereby enhancing the robustness of the final regression model. The \texttt{RNC} model performs best on the in-domain test but worse than the \texttt{SimCLR} model on the cross-domain tests. This is because it relies heavily on the ordering of pseudo-labels, which can be very noisy.

With our weakly supervised contrastive learning method (\texttt{Proposed} $\mathcal{L}_\text{coarse}$), we further improve performance over \texttt{SimCLR}. Our method generally enhances results on the cross-domain sets while maintaining comparable performance on the in-domain test; it achieves the best or the second-best scores for all cross-domain tests except EasyCall. For the EasyCall dataset, it performs slightly worse than \texttt{SimCLR}, but still improves SRCC by 0.23 compared to the \texttt{Baseline} model. Interestingly, with our proposed fine-grained supervised representation learning ($\mathcal{L}_\text{dist}$ and $\mathcal{L}_\text{con}$), performance on the in-domain test slightly improves; however, cross-domain performance generally degrades compared to the \texttt{Baseline}. We suggest that providing strong and fine-grained guidance via pseudo-labeling may harm generalizability by increasing overfitting to the in-domain set. However, when we increase $\tau$ for both methods (Figure~\ref{fig:tau_improvement}), the average improvement on the cross-domain tests generally increases, highlighting the effectiveness of weak supervision. 
% We further discuss the impact of $\tau$ in the ablation study.

\begin{figure*}[t]
     \centering
     \begin{subfigure}[b]{0.158\linewidth}
         \centering
         \includegraphics[width=\textwidth]{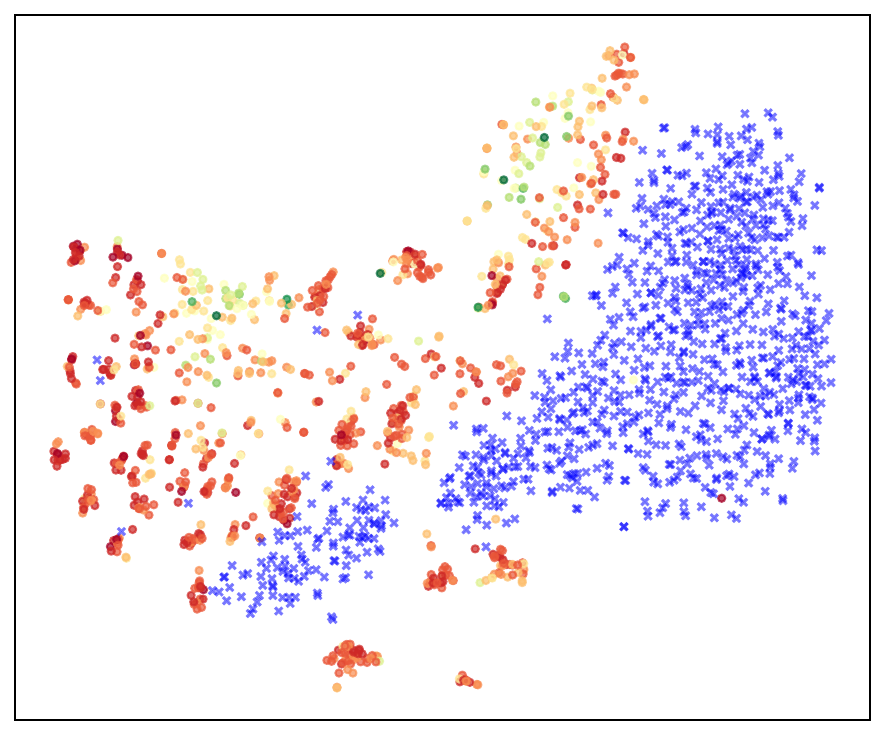}
         % \vspace{-0.05in}
         \caption{No contrast}
         \label{fig:embedding:random}
     \end{subfigure}
     % \qquad
     \begin{subfigure}[b]{0.158\linewidth}
         \centering
         \includegraphics[width=\textwidth]{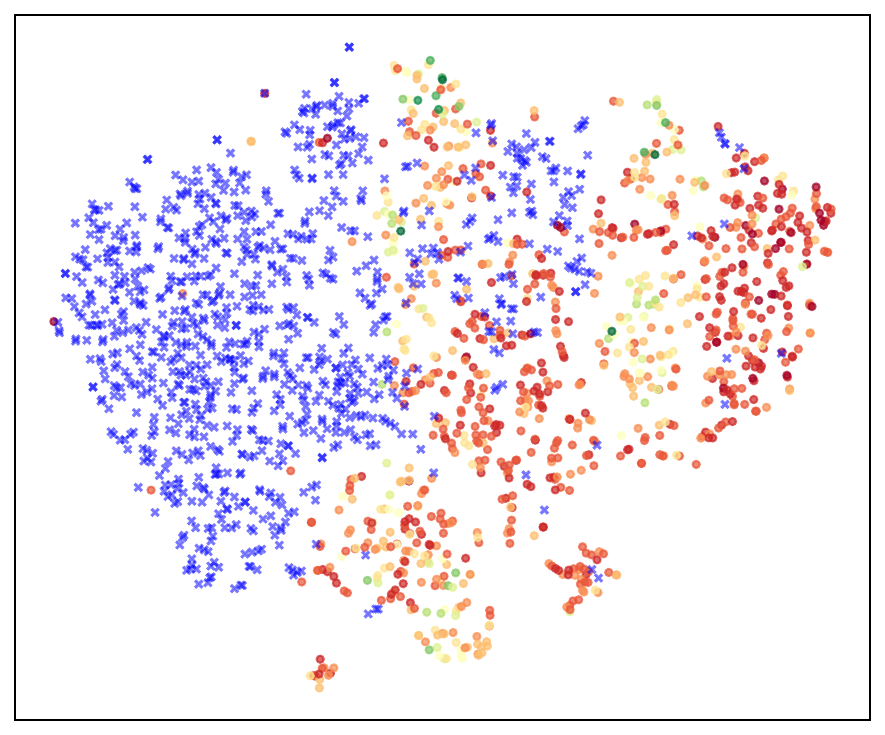}
         % \vspace{-0.05in}
         \caption{SimCLR}
         \label{fig:embedding:simclr}
     \end{subfigure}
     \begin{subfigure}[b]{0.158\linewidth}
         \centering
         \includegraphics[width=\textwidth]{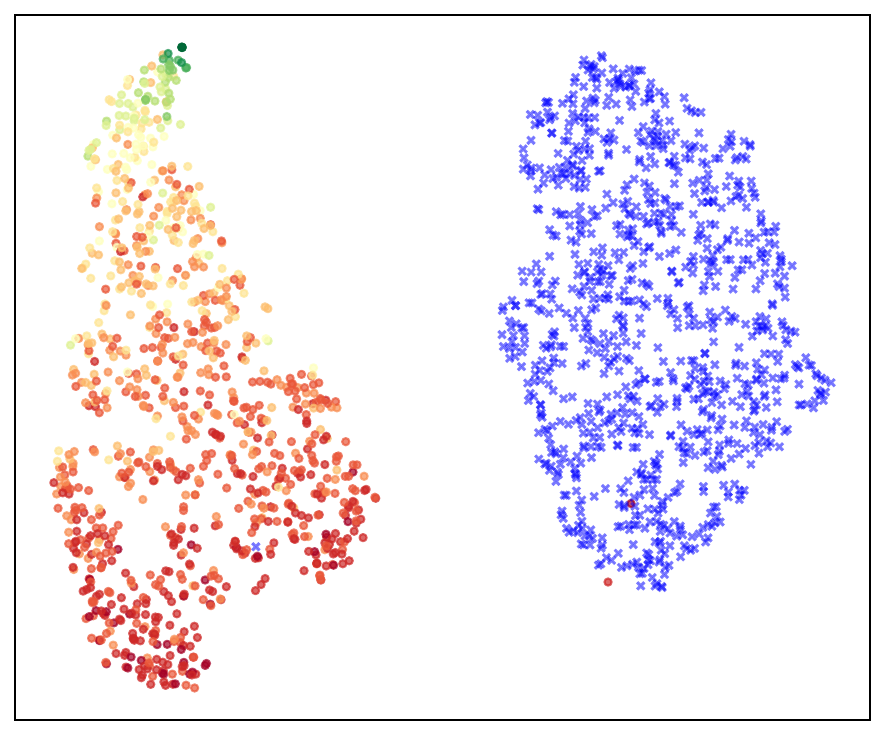}
         % \vspace{-0.05in}
         \caption{Rank-N-Contrast}
         \label{fig:embedding:rnc}
     \end{subfigure}
     % \qquad
     \begin{subfigure}[b]{0.158\linewidth}
         \centering
         \includegraphics[width=\textwidth]{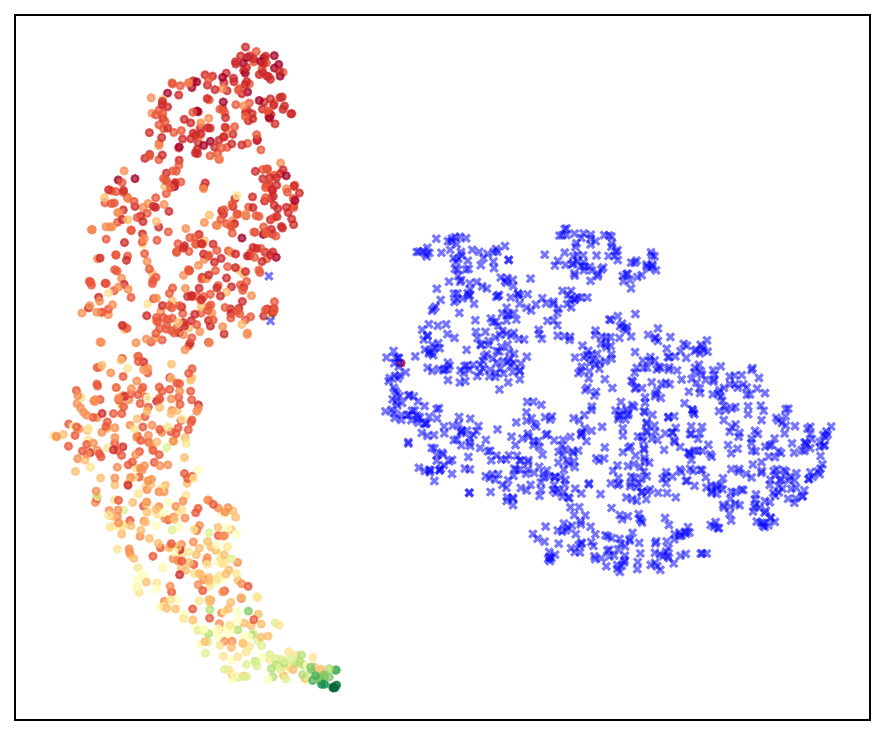}
         % \vspace{-0.05in}
         \caption{Proposed ($\mathcal{L}_\text{dis}$)}
         \label{fig:embedding:dis}
     \end{subfigure}
     \begin{subfigure}[b]{0.158\linewidth}
         \centering
         \includegraphics[width=\textwidth]{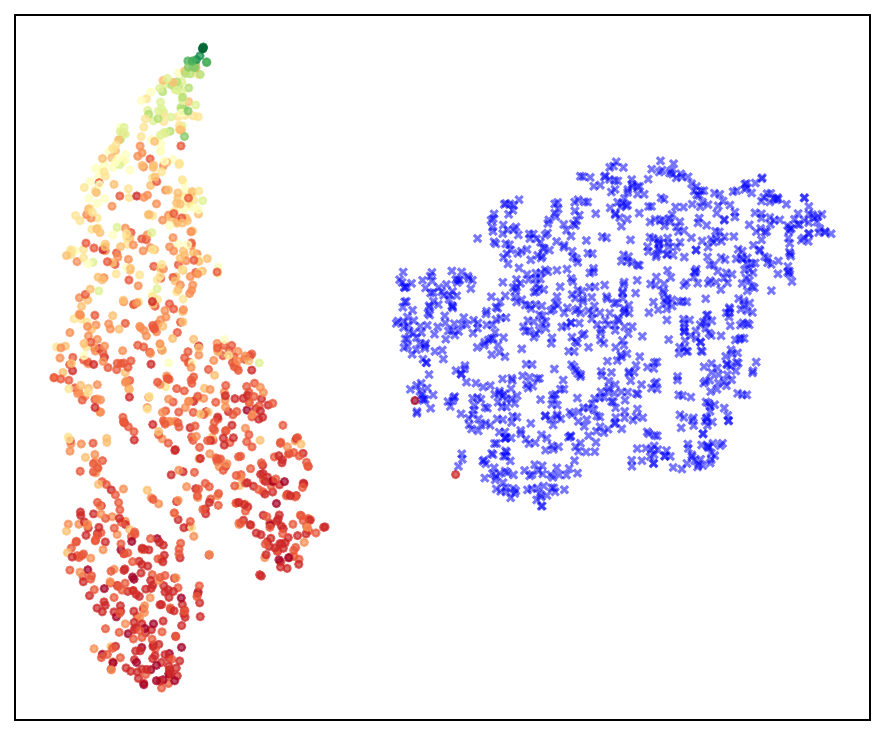}
         % \vspace{-0.05in}
         \caption{Proposed ($\mathcal{L}_\text{con}$)}
         \label{fig:embedding:con}
     \end{subfigure}
     % \qquad
     \begin{subfigure}[b]{0.175\linewidth}
         \centering
         \includegraphics[width=\textwidth]{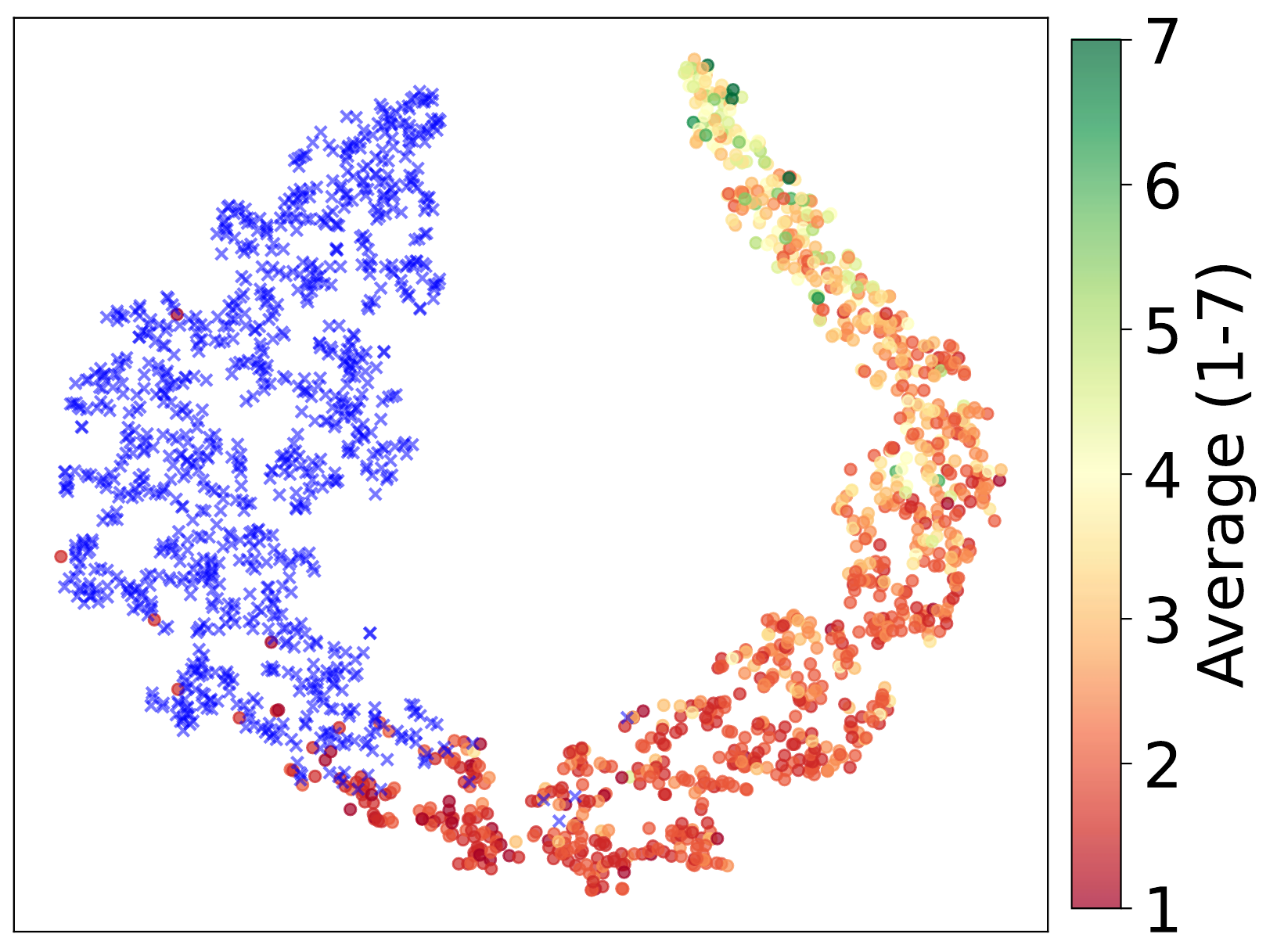}
         % \vspace{-0.05in}
         \caption{Proposed ($\mathcal{L}_\text{coarse}$)}
         \label{fig:embedding:coarse}
     \end{subfigure}
     % \qquad
     % \begin{subfigure}[b]{0.176\textwidth}
     %     \centering
     %     \includegraphics[width=\textwidth]{fig/p(zbar).pdf}\vspace{-0.05in}
     %     \caption{}
     %     \label{fig:augmented}
     % \end{subfigure}
        \caption{t-SNE figures after stage 2 with various contrastive loss choices. We randomly select 1000 samples in LibriSpeech and SAP training data. These are the embeddings right after the temporal pooling. Blue cross represents the LibriSpeech data, and circles indicate the SAP dataset. From red to green, the color indicates the low-severity to high-severity levels. Best viewed in color.}
        \vspace{\vspaceval}
        \label{fig:embedding}
\end{figure*}
The t-SNE \cite{tsne} plots of the embedding space after the temporal pooling layers are shown in Figure~\ref{fig:embedding}. Even without contrastive learning, the feature space exhibits a certain structure due to the powerful Whisper-large speech foundation model (Figure~\ref{fig:embedding:random}). However, embeddings from the SAP dataset form subclusters and occupy a separate region from the LibriSpeech embeddings, likely due to other task-irrelevant speech attributes such as speaker identity, limiting the robustness of the model. Without supervision (Figure~\ref{fig:embedding:simclr}), the embedding space becomes smoother and more harmonized with LibriSpeech, but still does not form a structure aligned with severity levels. In contrast, Figure~\ref{fig:embedding:rnc}, \ref{fig:embedding:dis}, \ref{fig:embedding:con}, and \ref{fig:embedding:coarse} show a smooth embedding space where SAP features are ordered based more on the target labels, resulting in a more informative representation for our downstream task. More importantly, Figure~\ref{fig:embedding:coarse} presents the smoothest transition from the LibriSpeech portion (blue crosses) to the level 1 examples in SAP, which explains the overall superior performance of \texttt{Proposed} $\mathcal{L}_\text{coarse}$ to others. This demonstrates the effectiveness of our representation learning. Meanwhile, we note that, with \texttt{RNC} and fine-grained supervision, the embedding spaces of the SAP and LibriSpeech datasets are separate, limiting the models' generalizability.

\subsection{Ablation Studies}
\label{sec:ablation}

\begin{table*}[t]
\caption{Ablation study of each stage of the Proposed ($\mathcal{L}_\text{coarse}$) model.}
\vspace{\tablecaptionvspaceval}
\centering
% \small
\resizebox{\linewidth}{!}{
\scriptsize
\renewcommand{\arraystretch}{1.2}
\begin{tabular}{c|ccc|cc|cc|cc|cc|cc|cc|cc}
\toprule
\multirow{3}{*}{\textbf{Index}} & \multicolumn{3}{c|}{\multirow{2}{*}{\textbf{Stage}}} & \multicolumn{2}{c|}{\textbf{In-domain}} 
& \multicolumn{12}{c}{\textbf{Cross-domain}} \\
% \cmidrule(lr){2-7} \cmidrule(lr){8-19}
\cline{5-18}
~ & ~ & ~ & ~ & \multicolumn{2}{c|}{SAP} 
& \multicolumn{2}{c|}{UASpeech}
& \multicolumn{2}{c|}{DysArinVox}
& \multicolumn{2}{c|}{EasyCall}
& \multicolumn{2}{c|}{EWA-DB}
& \multicolumn{2}{c|}{NeuroVoz}
& \multicolumn{2}{c}{Average}
\\
\cline{2-18}
& 1 & 2 & 3
& SRCC & PCC
& SRCC & PCC
& SRCC & PCC
& SRCC & PCC
& SRCC & PCC
& SRCC & PCC
& SRCC & PCC\\
\hline
1 & ~ & ~ & \cmark & \textbf{0.719} & \textbf{0.755} & 0.949 & 0.963 & 0.578 & 0.619 & 0.849 & 0.783 & \underline{0.709} & \underline{0.686} & 0.575 & 0.601 & 0.732 & 0.730 \\
2 & \cmark & ~ & \cmark & \textbf{0.719} & \textbf{0.755} & 0.949 & 0.963 & 0.577 & 0.620 & 0.856 & 0.785 & \underline{0.709} & \underline{0.686} & 0.571 & 0.598 & 0.732 & 0.730\\
3 & ~ & \cmark & \cmark & 0.681 & 0.740 & \underline{0.960} & \underline{0.968} & \underline{0.604} & \textbf{0.654} & \textbf{0.902} & \textbf{0.850} & 0.670 & 0.667 & \underline{0.596} & \underline{0.621} & \underline{0.746} & \textbf{0.752} \\
\hline
4 & \cmark & \cmark & \cmark & \underline{0.716} & \underline{0.750} & \textbf{0.975} & \textbf{0.976} & \textbf{0.631} & \underline{0.632} & \underline{0.872} & \underline{0.810} & \textbf{0.711} & \textbf{0.695} & \textbf{0.617} & \textbf{0.630} & \textbf{0.761} & \underline{0.749} \\
% \hline
\bottomrule
\end{tabular}
}
\label{tab:results-stage-albation}
\end{table*}

\subsubsection{Effectiveness of each stage} 
\label{sec:ablation:stage}
Table~\ref{tab:results-stage-albation} shows the ablation results for each stage of \texttt{Proposed} $\mathcal{L}_\text{coarse}$. The first and second rows are nearly identical; without the contrastive learning stage (Stage 2), the model cannot benefit from the additional pseudo-labeled dataset, as the pseudo-labels essentially imitate the distribution of the SAP-labeled dataset. Conversely, without the pseudo-labeling stage (third row), we have no way but to label the LibriSpeech portion as non-dysarthric, while all the unlabeled SAP data are considered dysarthric, instead of using a threshold $\beta$ as in eq. \eqref{eq:corse}. This prevents the model from distinguishing speech with low severity levels (i.e., speech close to typical) within the SAP unlabeled dataset, resulting in incorrect supervision of the embedding space and causing misalignment between the SAP and LibriSpeech datasets.

\begin{table*}[t]
\caption{Ablation study on additional LibriSpeech dataset, unlabeled portion of SAP dataset, and variance regularization objective.}
\vspace{\tablecaptionvspaceval}
\centering
% \small
% \resizebox{0.8\textwidth}{!}{
% \footnotesize
\scriptsize
\renewcommand{\arraystretch}{1.2}
\begin{tabular}{l|cc|cc|cc|cc|cc|cc|cc}
\toprule
\multirow{3}{*}{\textbf{Model}} & \multicolumn{2}{c|}{\textbf{In-domain}} 
& \multicolumn{12}{c}{\textbf{Cross-domain}} \\
% \cmidrule(lr){2-7} \cmidrule(lr){8-19}
\cline{2-15}
~ & \multicolumn{2}{c|}{SAP} 
& \multicolumn{2}{c|}{UASpeech}
& \multicolumn{2}{c|}{DysArinVox}
& \multicolumn{2}{c|}{EasyCall}
& \multicolumn{2}{c|}{EWA-DB}
& \multicolumn{2}{c|}{NeuroVoz}
& \multicolumn{2}{c}{Average}
\\
\cline{2-15}
& SRCC & PCC
& SRCC & PCC
& SRCC & PCC
& SRCC & PCC
& SRCC & PCC
& SRCC & PCC
& SRCC & PCC\\
\hline
Proposed ($\mathcal{L}_{\text{coarse}}$) & \textbf{0.716} & 0.750 & \textbf{0.975} & \underline{0.976} & \underline{0.631} & \underline{0.632} & \underline{0.872} & \underline{0.810} & \textbf{0.711} & \textbf{0.695} & \underline{0.617} & \underline{0.630} & \textbf{0.761} & \textbf{0.749} \\
% \hline
\hline
w/o $\mathcal{D}_{\text{Libri}}$ & 0.706 & 0.743 & 0.952 & 0.964 & 0.605 & 0.630 & 0.832 & 0.746 & \underline{0.692} & \underline{0.668} & 0.590 & 0.613 & 0.734 & 0.724\\
w/o $\mathcal{D}_{\text{unlabeled}}$ & \underline{0.715} & \textbf{0.755} & 0.950 & 0.956 & 0.609 & 0.625 & \textbf{0.877} & 0.799 & 0.681 & 0.660 & \textbf{0.632} & \textbf{0.659} & 0.750 & 0.740\\
w/o $\mathcal{L}_{\text{var}}$ & 0.711 & \underline{0.753} & \underline{0.963} & \textbf{0.978} & \textbf{0.661} & \textbf{0.658} & 0.862 & \textbf{0.813} & 0.683 & 0.660 & 0.606 & 0.616 & \underline{0.755} & \underline{0.745}\\
% \hline
% Naturalness as label & 0.722 & 0.746 & 0.962 & 0.965 & 0.586 & 0.648 & 0.834 & 0.811 & 0.724 & 0.716 & 0.658 & 0.664 & 0.753 & 0.761 \\
% Intelligibility as label &\\
\bottomrule
\end{tabular}
% }
% \vspace{\vspaceval}
\label{tab:results-other-ablation}
\end{table*}

\subsubsection{Effectiveness of $\mathcal{D}_\text{Libri}$, $\mathcal{D}_\text{unlabeled}$, and $\mathcal{L}_\text{var}$}
\label{sec:ablation:data_loss}
In Table~\ref{tab:results-other-ablation}, we observe that $\mathcal{D}_\text{Libri}$ plays a crucial role in enhancing model performance. Although performance increases on some cross-domain tests, the average SRCC score remains almost the same as that of the \texttt{Baseline} model. This demonstrates that incorporating additional typical speech with diverse speakers and acoustic conditions is effective for improving the robustness of dysarthric speech severity prediction. Excluding $\mathcal{D}_{\text{unlabeled}}$ also degrades performance, but less so than excluding $\mathcal{D}_{\text{Libri}}$, because $\mathcal{D}_{\text{unlabeled}}$ and $\mathcal{D}_{\text{labeled}}$ share overlapping speakers, meaning the model retains partial exposure to those speakers through $\mathcal{D}_{\text{labeled}}$ alone. Without $\mathcal{L}_\text{var}$, the overall performance degrades, highlighting its effectiveness in improving robustness in representation learning.

\subsubsection{Temperature value $\tau$}
\label{sec:ablation:tau}
\begin{figure}[t]
    \centering
     \begin{subfigure}[b]{\linewidth}
         \centering
         \includegraphics[width=\textwidth]{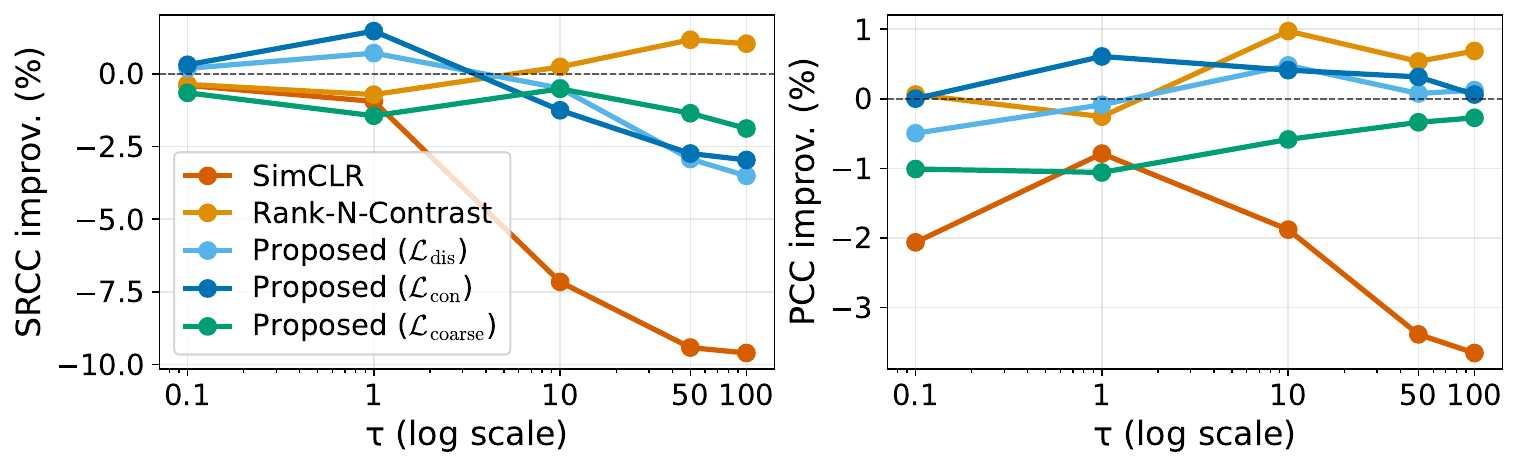}
         % \vspace{-0.05in}
         \caption{In-domain (SAP dataset)}
         \label{fig:input_dist}
     \end{subfigure}
     % \qquad
     \begin{subfigure}[b]{\linewidth}
         \centering
         \includegraphics[width=\textwidth]{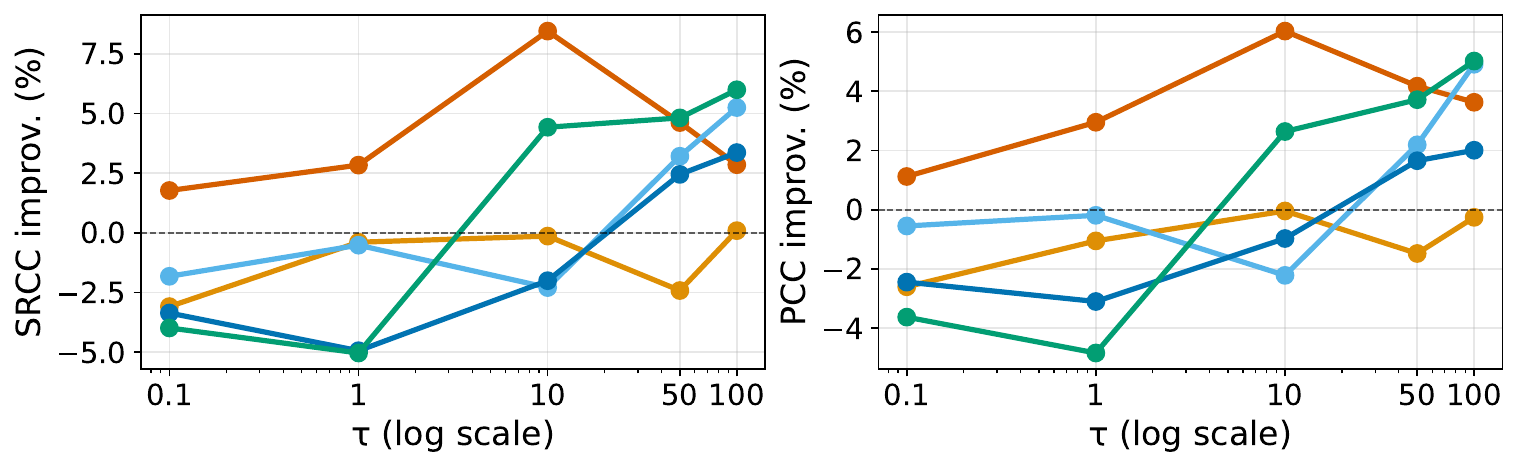}
         % \vspace{-0.05in}
         \caption{Cross-domain (average)}
         \label{fig:feature_dist}
     \end{subfigure}
     
    % \includegraphics[width=1.0\linewidth]{figures/graph_by_tau_improvement.pdf}
    % \vspace{-0.05in}
    \caption{The improvement percentages of SRCC and PCC over the \texttt{Baseline} model vary with different values of $\tau$. (a) In-domain testset (SAP dataset) and (b) average scores of cross-domain testsets. In general, the performance of our proposed methods improves as $\tau$ increases. Although \texttt{SimCLR} achieves the best cross-domain average performance at $\tau=10$, its in-domain test performance deteriorates significantly, highlighting the robustness of our proposed methods.}
    \vspace{\vspaceval}
    \label{fig:tau_improvement}
\end{figure}

The SRCC and PCC improvements over the \texttt{Baseline} model are shown in Figure~\ref{fig:tau_improvement}. For the \texttt{Proposed} models with different contrastive losses, performance generally improves as the temperature parameter $\tau$ increases. This highlights the importance of weak supervision in contrastive learning for harmonizing the three training datasets: $\mathcal{D}_{\text{labeled}}$, $\mathcal{D}_{\text{pseudo}}$, and $\mathcal{D}_{\text{Libri}}$. While \texttt{SimCLR} achieves the best cross-domain performance at $\tau=10$, its in-domain performance drops significantly, indicating reduced stability compared to the proposed methods.

\begin{figure}[t]
     \centering
     \begin{subfigure}[b]{0.4\linewidth}
         \centering
         \includegraphics[width=\textwidth]{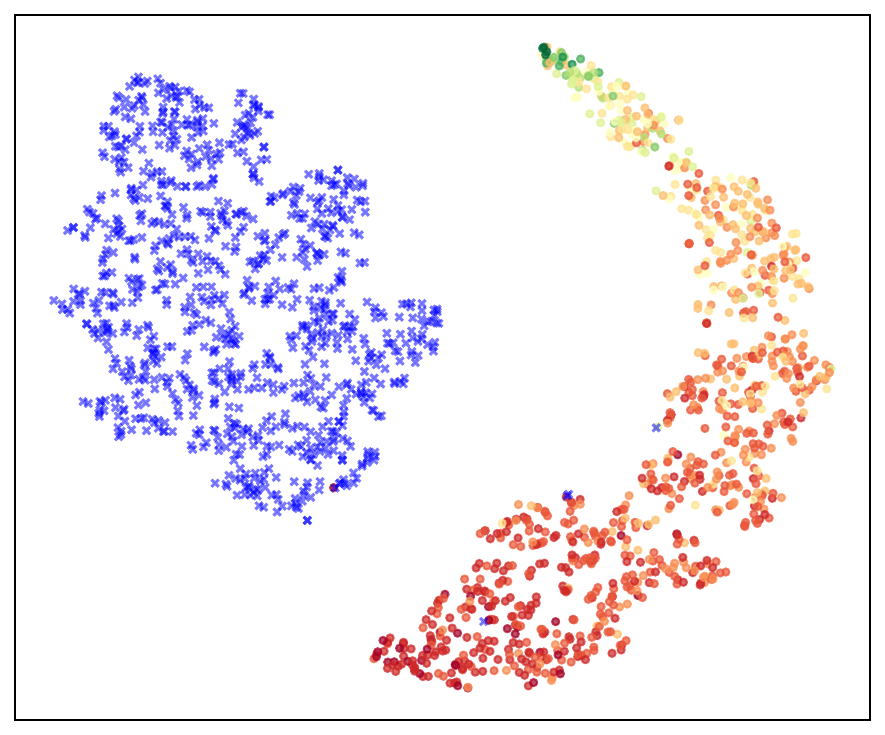}\vspace{-0.05in}
         \caption{$\tau=0.1$}
         \label{fig:input_dist}
     \end{subfigure}
     % \qquad
     \begin{subfigure}[b]{0.4\linewidth}
         \centering
         \includegraphics[width=\textwidth]{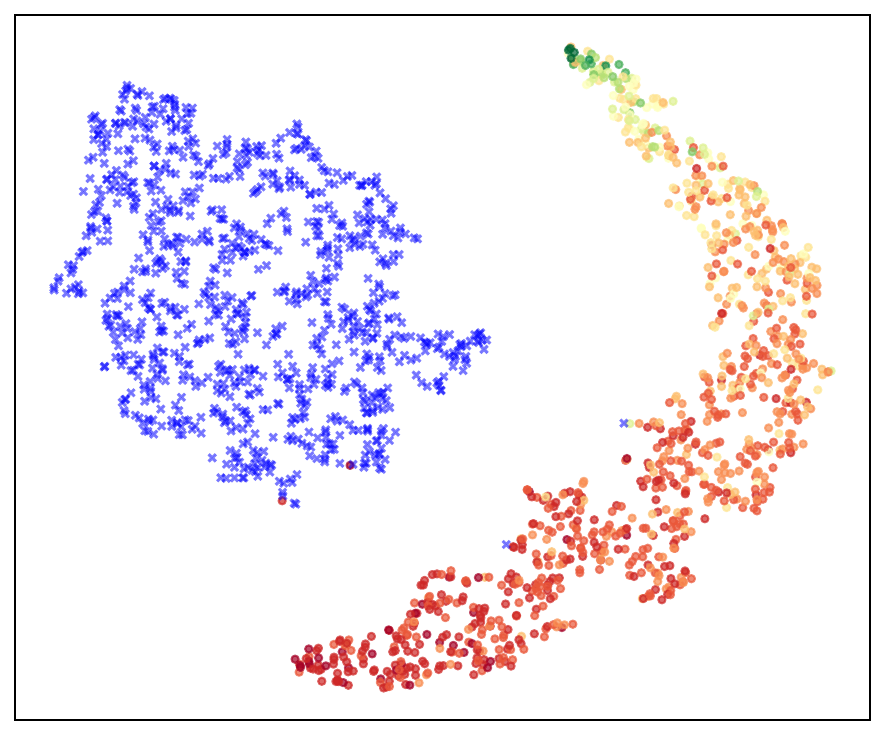}\vspace{-0.05in}
         \caption{$\tau=1.0$}
         \label{fig:feature_dist}
     \end{subfigure}
     % \qquad
     \begin{subfigure}[b]{0.4\linewidth}
         \centering
         \includegraphics[width=\textwidth]{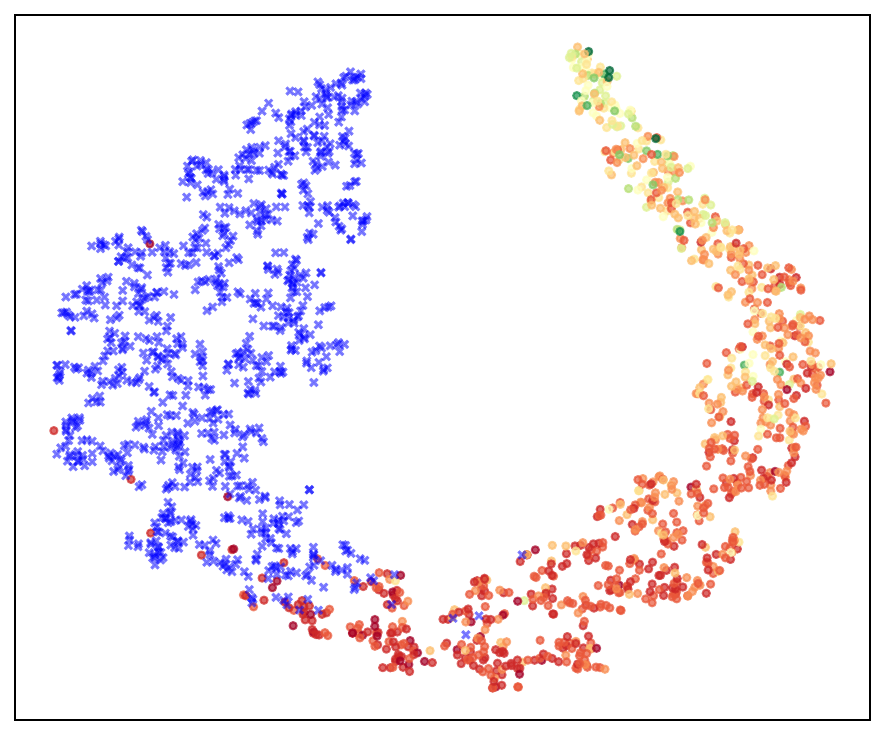}\vspace{-0.05in}
         \caption{$\tau=10.0$}
         \label{fig:low-resource-samples}
     \end{subfigure}
     \begin{subfigure}[b]{0.4\linewidth}
         \centering
         \includegraphics[width=\textwidth]{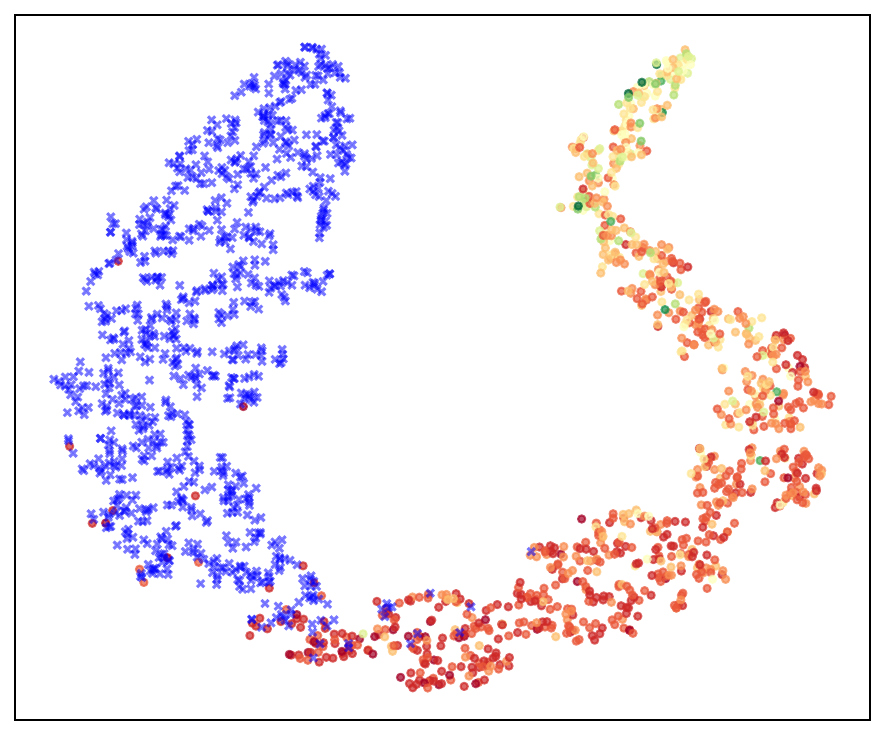}\vspace{-0.05in}
         \caption{$\tau=100.0$}
         \label{fig:low-resource-samples}
     \end{subfigure}
     % \qquad
     % \begin{subfigure}[b]{0.176\textwidth}
     %     \centering
     %     \includegraphics[width=\textwidth]{fig/p(zbar).pdf}\vspace{-0.05in}
     %     \caption{}
     %     \label{fig:augmented}
     % \end{subfigure}
        \caption{Embedding spaces with different $\tau$. Since the LibriSpeech and SAP datasets have distinct characteristics, they can be considered easy positive/negative pairs. Therefore, with a small $\tau$, contrastive learning tends not to associate pairs from the LibriSpeech and SAP datasets. In contrast, with a large $\tau$, they are better harmonized, enhancing the robustness of the downstream regression model.}
        \vspace{\vspaceval}
        \label{fig:embedding_tau}
\end{figure}

When $\tau$ is small, the contrastive loss emphasizes hard positive and negative pairs. Since SAP and LibriSpeech may differ significantly in their acoustic characteristics, they are easily distinguishable. As a result, a small $\tau$ encourages the model to focus on dataset-specific structures rather than learning shared representations across datasets. This effect is also reflected in Figure~\ref{fig:embedding_tau}: with a small $\tau$, SAP and LibriSpeech representations are clearly separated in the t-SNE visualization, whereas larger $\tau$ values produce more aligned and harmonized embeddings.

These observations suggest that using a larger $\tau$ facilitates better integration of the external typical dataset with dysarthric speech data, ultimately improving the robustness of the model.

\section{Broad Impact}
This work advances scalable and automated assessment of dysarthric speech severity, with potential benefits for clinical monitoring, rehabilitation, and the development of more inclusive speech technologies. Our experiments further suggest that the proposed approach can generalize to non-English languages, where labeled data are often even more scarce than in English. Moreover, while our study focuses on dysarthric speech assessment, the proposed framework for leveraging weak supervision and pseudo-labeled data may also be applicable to other domains where labeled data are limited, enabling scalable data augmentation and representation learning beyond speech tasks. While we believe this tool has the potential to positively impact accessibility for individuals with dysarthria, several considerations remain important. Automated predictions should not be interpreted as clinical diagnoses. In addition, as dysarthric speech data are closely tied to health conditions, privacy protection and responsible data use are essential in any real-world deployment. Appropriate safeguards are necessary to ensure ethical use and to prevent misuse in sensitive decision-making contexts.

% This work advances scalable and automated assessment of dysarthric speech severity, with potential benefits for clinical monitoring, rehabilitation, and the development of more inclusive speech technologies. 
% % We improve the robustness of such automated tools by leveraging largely unlabeled dysarthric speech alongside typical speech data. 
% Our experiments further suggest that the proposed approach can generalize to non-English languages, where labeled data are often even more scarce than in English.
% While we believe this tool has the potential to positively impact accessibility for individuals with dysarthria, several considerations remain important. Automated predictions should not be interpreted as clinical diagnoses. Moreover, as dysarthric speech data are closely tied to health conditions, privacy protection and responsible data use are essential in any real-world deployment, and appropriate safeguards are necessary to ensure ethical use and to prevent misuse in sensitive decision-making contexts. \minje{It might be nice if we can argue that the algorithms proposed in the paper can be used for non-speech darta augmentation, too. }

\section{Conclusion and Future Works}
In this work, we proposed a three-stage framework for robust dysarthric speech severity estimation that leverages unlabeled dysarthric speech and large-scale typical speech through pseudo-labeling and weakly supervised contrastive learning. By adapting Whisper representations toward task-relevant severity structure, our approach improves generalization across different etiologies, languages, and labeling schemes. Experimental results demonstrate strong cross-domain robustness, achieving an average SRCC of 0.761 on unseen datasets while maintaining in-domain performance; 0.415 and 0.311 point improvement compared to UTMOS and SpICE, respectively. Our ablation studies further show that weak supervision via coarse grouping in contrastive learning, along with a higher temperature ($\tau$), plays a critical role in improving generalization.

In this work, training relied on a single dysarthric speech dataset, SAP. Although SAP is relatively large-scale, incorporating multilingual dysarthric datasets during training could further improve robustness. In addition, we showed that the learned representation space has the potential to serve as an interpretable severity dimension. Further analysis of this space to extract meaningful insights and understand the factors influencing predictions would be an important direction for future work.
Finally, we plan to use the proposed severity predictor as an automatic labeling tool for unlabeled dysarthric speech and apply it to support downstream systems such as dysarthric ASR and speech enhancement.

\newpage

\section{Acknowledgments}

% {\color{blue}Acknowledgments should be included only in the camera-ready version, not in the version submitted for review. For regular papers, pages 5 and 6, and for long papers, pages 9 and 10, are reserved exclusively for acknowledgments, disclosures of the use of generative AI tools, and references. No other content may appear on these pages. Any appendices must be contained within the first four pages for regular papers and within the first eight pages for long papers.

% Acknowledgments and references may begin on an earlier page if space permits.}

% \ifcameraready
%      The Interspeech 2026 organizers
% \else
%      The authors
% \fi
% This work was supported by Electronics and Telecommunications Research Institute (ETRI) grant funded by the Korean government [26ZC1100, Development of Spatial Media Technology and Interaction Technology for Convergence of the Real and Virtual World].

This work was supported in part by the Institute of Information \& communications Technology Planning \& evaluation (IITP) grant funded by the Korean government (MSIT) [No. RS-2022-II220184, Development and Study of AI Technologies to Inexpensively Conform to Evolving Policy on Ethics] as well as the National Science Foundation under Grant No. 2512987.

\section{Generative AI Use Disclosure}
The authors acknowledge the use of an AI tool for copyediting and polishing the English language in this manuscript. The tool was used only to improve clarity, grammar, and style, and was not used to generate substantial portions of the manuscript or to develop the scientific content. All research design, experiments, analyses, and conclusions were conducted by the authors, who take full responsibility for the content of the paper.

\bibliographystyle{IEEEtran}
\bibliography{conference_list,mybib_clean}

\end{document}